\documentclass[aps,prd,twocolumn,superscriptaddress,nofootinbib]{revtex4}
\usepackage{graphicx}
\usepackage{dcolumn}
\usepackage{amssymb,amsmath,amsthm,mathrsfs}
\usepackage{epstopdf}
\usepackage{lipsum}
\usepackage{hyperref}
\usepackage{enumerate}
\usepackage[makeroom]{cancel}

\begin{document}
    
    \newcommand{\Eq}[1]{Eq. \ref{eqn:#1}}
    \newcommand{\Fig}[1]{Fig. \ref{fig:#1}}
    \newcommand{\Sec}[1]{Sec. \ref{sec:#1}}
    \newcommand{\PHI}{\phi}
    \newcommand{\vect}[1]{\mathbf{#1}}
    \newcommand{\Del}{\nabla}
    \newcommand{\unit}[1]{\mathrm{#1}}
    \newcommand{\x}{\vect{x}}
    \newcommand{\ScS}{\scriptstyle}
    \newcommand{\ScScS}{\scriptscriptstyle}
    \newcommand{\xplus}[1]{\vect{x}\!\ScScS{+}\!\ScS\vect{#1}}
    \newcommand{\xminus}[1]{\vect{x}\!\ScScS{-}\!\ScS\vect{#1}}
    \newcommand{\diff}{\mathrm{d}}
         \newcommand{\keq}{k_{\rm eq}}
    
    \newcommand{\be}{\begin{equation}}
        \newcommand{\ee}{\end{equation}}
    \newcommand{\bea}{\begin{eqnarray}}
        \newcommand{\eea}{\end{eqnarray}}
    \newcommand{\vu}{{\mathbf u}}
    \newcommand{\ve}{{\mathbf e}}
    \newcommand{\vk}{{\mathbf k}}
    \newcommand{\vx}{{\mathbf x}}
    \newcommand{\vy}{{\mathbf y}}
    \newcommand{\bx}{{\bf x}}
    \newcommand{\bk}{{\bf k}}
    \newcommand{\br}{{\bf r}}
    
    \newcommand{\uden}{\underset{\widetilde{}}}
    \newcommand{\den}{\overset{\widetilde{}}}
    \newcommand{\denep}{\underset{\widetilde{}}{\epsilon}}
    
    \newcommand{\nn}{\nonumber \\}
    \newcommand{\dd}{\diff}
    \newcommand{\fr}{\frac}
    \newcommand{\del}{\partial}
    \newcommand{\eps}{\epsilon}
    \newcommand\CS{\mathcal{C}}
    
    \def\be{\begin{equation}}
        \def\ee{\end{equation}}
    \def\ben{\begin{equation*}}
        \def\een{\end{equation*}}
    \def\bea{\begin{eqnarray}}
        \def\eea{\end{eqnarray}}
    \def\bal{\begin{align}}
        \def\eal{\end{align}}
    
    \def\TT{{\rm TT}}
    \def\GW{{_{_{\rm GW}}}}
        \def\RD{{\rm RD}}
        \def\KD{{_{\rm KD}}}
            \def\MD{{\rm MD}}

\title{Gravitational wave production from the decay\\
of the Standard Model Higgs field after inflation}

\newcommand{\addressIFT}{Instituto de F\'isica Te\'orica UAM-CSIC, Universidad Auton\'oma de Madrid, Cantoblanco, 28049 Madrid, Spain.}
\newcommand{\addressCERN}{CERN, Theory Division, 1211 Geneva, Switzerland.}
\newcommand{\addressGeneva}{D\'epartement de Physique Th\'eorique and Center for Astroparticle Physics, Universit\'e de Gen\`eve, 24 quai Ernest Ansermet, CH--1211 Gen\`eve 4, Switzerland.}

\author{Daniel G. Figueroa}
\affiliation{\addressCERN}

\author{Juan Garc\'ia-Bellido}
\affiliation{\addressIFT}

\author{Francisco Torrent\'i\,}
\affiliation{\addressIFT}

\date{\today}

\begin{abstract}
During or towards the end of inflation, the Standard Model (SM) Higgs field forms a condensate with a large amplitude. Following inflation, the condensate oscillates, decaying non-perturbatively into the rest of the SM species. The resulting out-of-equilibrium dynamics converts a fraction of the energy available into gravitational waves (GW). We study this process using classical lattice simulations in an expanding box,
following the energetically dominant electroweak gauge bosons {\small$W^\pm$} and {\small$Z$}. We characterize the GW spectrum as a function of the running couplings, Higgs initial amplitude, and post-inflationary expansion rate. As long as the SM is decoupled from the inflationary sector, the generation of this background is universally expected, independently of the nature of inflation. Our study demonstrates the efficiency of GW emission by gauge fields undergoing parametric resonance. The initial energy of the Higgs condensate represents however, only a tiny fraction of the inflationary energy. Consequently, the resulting background is very suppressed, with an amplitude {\small$h^2 \Omega_{\rm GW}^{(o)} \lesssim 10^{-29}$} today. The amplitude can be boosted to {\small$h^2 \Omega_{\rm GW}^{(o)} \lesssim 10^{-16}$}, if following inflation the universe undergoes a kination-domination stage; however the background is shifted in this case to high frequencies {\small$f_p \lesssim 10^{11} {\rm Hz}$}. In all cases the signal is out of the range of current or planned GW detectors. This background will therefore remain, most likely, as a curiosity of the SM.
\end{abstract}

\maketitle

\section{Introduction}
\label{sec:Intro}

Gravitational waves (GW) are ripples of the spacetime which propagate at the speed of light. Until very recently, the existence of GW had only been proven indirectly, through the modulation of the orbital period of binary pulsars \cite{HulseTaylor:1975a}.  Advanced LIGO~\cite{GWdetection:abc} has just announced, however, the first direct detection of GW from the coalescence of two massive black holes. This historical event opens a new window into the Universe, which will allow us to probe astrophysical and cosmological environments previously inaccessible. This milestone detection will very likely inaugurate a new era in cosmology.

The Universe is presumed to be permeated by various GW backgrounds of cosmological origin. From inflation, we expect an almost scale-invariant background~\cite{StarobinskyGW1979}. From non-equilibrium phenomena after inflation, we expect a strong production of GW from (p)reheating~\cite{GWs1997KhlebnikovTkachev, GarciaBellido:1998gm, GWs2006EastherLim, GWsPRL2007FigueroaGarciaBellido, GWsPRD2007FigueroaGarciaBellidoSastre, GWs2007DufauxEtAl, GWs2009HybridInfDufauxEtAl}, phase transitions~\cite{Kamionkowski:1993fg,Caprini:2007xq,Huber:2008hg,Hindmarsh:2013xza, Hindmarsh:2015qta}, or cosmic defects~\cite{GWsAbelianHiggsHybridPreheatingDufauxEtAl2010, GWsGeneral2004DamourVilenkin, Vachaspati:1984gt, JonesSmith:2007ne, GWsSelfOrderingScalarFieldsFenuEtAl2009, GWsUniversalBackgroundFromCDsFigueroaEtAl2012}. A direct detection of these backgrounds will open a new window into the very early universe, probing physical phenomena at energies beyond the reach of particle colliders~\cite{Maggiore2000}.
The development of GW detectors like Advanced VIRGO~\cite{Acernese:2015gua}, Advanced LIGO~\cite{Harry:2010zz}, KAGRA~\cite{Somiya:2011np}, and eLISA~\cite{AmaroSeoane:2012km}, aims to make this possible in the near future. We need therefore to characterize all possible signals in order to better understand a future detection.

In this work we study the production of GW within the framework of the Standard Model (SM) of particle physics. The discovery of the Higgs boson at the Large Hadron Collider (LHC)~\cite{ATLAS2012,CMS2012} has triggered an intense work to analyze its possible cosmological consequences in the early universe~\cite{Espinosa:2007qp,Enqvist:2013kaa,Figueroa:2014aya,KariEtAlNonAbelian14,Figueroa:2015hda,Enqvist:2015nrw,Enqvist:2014bua,Hook:2014uia,Kobakhidze:2013tn, Kobakhidze:2014xda, Spencer-Smith:2014woa, Shkerin:2015exa, Espinosa:2015qea, Gross:2015bea, Branchina:2015nda, DiVita:2015bha, DiLuzio:2015iua, Grobov:2015ooa,Kamada:2015aqa}. Here we consider the non-perturbative decay of the SM Higgs condensate after inflation~\cite{Enqvist:2013kaa, Figueroa:2014aya, KariEtAlNonAbelian14, Figueroa:2015hda, Enqvist:2015nrw}, assuming that the general features of the SM are valid all the way up to the inflationary scale. 

A compelling possibility is the \emph{Higgs-Inflation} scenario~\cite{Bezrukov:2007ep,Bezrukov:2010jz}, in which the SM Higgs is responsible for inflation, thanks to a large non-minimal coupling to gravity. Interestingly, even if inflation is driven by an inflaton scalar field other than the SM Higgs (typically a singlet under the SM), the Higgs may still play a very relevant role after inflation. In this work we assume that the SM Higgs is not responsible for inflation, and indeed we consider it very weakly coupled to the inflationary sector, even at loop order. Under this circumstance, we can distinguish two possibilities: $i)$ If the Higgs is minimally coupled to gravity, it behaves as a spectator field during inflation. It then forms a condensate with a large {\it vacuum expectation value} (VEV), and a correlation length exponentially larger than the Hubble radius~\cite{Starobinsky:1994bd, Kunimitsu:2012xx}. If on the contrary, $ii)$ the Higgs is non-minimally coupled to gravity with a sufficiently large coupling, the Higgs is not excited during inflation~\cite{Espinosa:2007qp,Herranen:2014cua}, but it is however strongly excited during the transition period 
at the end of inflation~\cite{Herranen:2015ima}. In this case, the Higgs forms also a condensate with large VEV, but with a correlation only of the order of the Hubble radius. 

In either case $i)$ or $ii)$, shortly after the end of inflation, the Higgs condensate starts oscillating around the minimum of its potential. This gives rise to its decay into all the species of the SM, as the latter are non-perturbatively excited through parametric effects~\cite{VaccaHiggs,Enqvist:2013kaa, Figueroa:2014aya, KariEtAlNonAbelian14, Figueroa:2015hda, Enqvist:2015nrw} (see also \cite{Bezrukov:2008ut, GarciaBellido:2008ab,Figueroa:2009jw} in the \emph{Higgs-Inflation} context). All the SM species coupled directly to the Higgs, i.e.~the electroweak gauge bosons {\small$W^\pm, Z$}, and the massive fermions (quarks and charged leptons), are all highly excited. This is a violent non-equilibrium process, creating large time-dependent matter density inhomogeneities, which therefore act as a classical source of GW. 

The decay of the Higgs into fermions, and the corresponding GW production, was studied in~\cite{Figueroa:2014aya}, following the formalism of~\cite{Enqvist:2012im,Figueroa:2013vif}. Fermions are excited through parametric effects~\cite{Greene:1998nh, Greene:2000ew}, though the growth of their occupation numbers is Pauli blocked. The most energetic fermion species excited is the top quark, since its Yukawa coupling is the largest one within the SM. In this paper we focus instead in the production of GW by the gauge bosons. The gauge field production is expected indeed to be more efficient than that of fermions, as their occupation numbers grow exponentially~\cite{Enqvist:2013kaa,KariEtAlNonAbelian14,Figueroa:2015hda,Enqvist:2015nrw}. 
Most of the energy of the Higgs condensate is actually transferred into the electroweak {\small$W^\pm, Z$} gauge bosons. Therefore, even if the final GW background is contributed by all the Higgs decay product species, the gauge fields we study here represent in fact the dominant contributors.

The decay of the SM Higgs condensate into gauge bosons after inflation, has been studied recently very extensively. It was first studied in~\cite{Enqvist:2013kaa,KariEtAlNonAbelian14} with analytical methods based on the linear regime,
and later extended in~\cite{Figueroa:2015hda}.
Beyond the linear regime, a full set of lattice simulations of the process was presented in~\cite{Figueroa:2015hda}, modeling the SM gauge interactions with an Abelian-Higgs set-up. Although this is just an approximation to the gauge structure of the electroweak interactions, the non-Abelian effects can be arguably neglected for a large fraction of the physically motivated values of the Higgs self-coupling. The outcome of these simulations, though neglecting the truly non-Abelian structure, represent a precise calculation of the dynamics of the SM after inflation, fully incorporating the nonlinear and non-perturbative effects of the SM gauge interactions between the Higgs and the {\small$W^\pm$} and {\small$Z$} gauge bosons. 

More recently, lattice simulations presented in~\cite{Enqvist:2015nrw} have considered the non-Abelian structure of the SM. 
They have shown interesting effects due to the new non-linearities introduced. The non-Abelian corrections are however suppressed by the smallness of the Higgs self-coupling~\cite{KariEtAlNonAbelian14,Figueroa:2015hda}. In high-energy inflationary models, the Higgs self-coupling runs in fact into small values~\cite{Degrassi:2012ry,Bezrukov:2012sa}, making the non-Abelian corrections less relevant, the larger the energy scale. In this paper we are mostly interested in scenarios with the highest possible energy scale of inflation, as this enhances the production of GW in the system. Therefore, the use of an Abelian modeling will suffice for our aim to study the GW production from the SM fields after inflation. 

The structure of this paper is as follows. In Section~\ref{sec:HiggsExcitation} we review the creation of a Higgs condensate during or towards the end of inflation. In Section~\ref{sec:PostInfDynamics} we review the post-inflationary dynamics of the Higgs and of its decay products, summarizing the results of~\cite{Figueroa:2015hda}. In Section~\ref{sec:GW} we discuss our formalism to study GW production in this process. In Section~\ref{sec:GWlatticeSimulations} we present our results, describing the general features of the GW spectra obtained from our lattice simulations. In Section \ref{sec:parametrization} we parametrize the {\rm GW} spectra as a function of the Higgs initial amplitude, Higgs self-coupling, and post-inflationary expansion rate. In section \ref{subsec:GWtoday} we discuss how the GW background redshifts until today. Finally, in Section~\ref{sec:Discussion} we wrap up our results and conclude. 

From now on, {\small$m_p^2 = (8 \pi G)^{-1} = 2.44\cdot10^{18}$} GeV is the reduced Planck mass, and we consider a
flat Friedman-Robertson-Walker background metric 
{\small$ds^2 =  a^2 (t) (-dt ^2 + d \vec{x}^2 )$}, with {\small$t$} the conformal time and {\small$a(t)$} the scale factor.

\section{Higgs excitation during (or towards the end of) inflation}
\label{sec:HiggsExcitation}

Relevant properties of the SM Higgs, like the nature of its gravitational coupling, or its coupling to the inflationary sector, 
are currently unknown. As a consequence, the role played by the Higgs, and in general its dynamics during the early Universe, are uncertain. In this paper we consider the Higgs to be sufficiently weakly coupled to the inflationary sector, so that the Higgs does not develop a super-Hubble mass during inflation. The need to reheat the Universe after inflation requires, however, the presence of a Higgs-inflaton coupling, induced through radiative corrections from some mediator field(s). 

The only scale-free renormalizable Higgs-inflaton operator is {\small$g^2\phi^2\Phi^\dag\Phi$}, where {\small$\Phi$} is the SM Higgs doublet, {\small$\phi$} is the inflaton, and {\small$g^2$} is the coupling strength. In order not to spoil the inflationary predictions, it is required that {\small$g^2 \lesssim g_*^2 = 10^{-6}$}~\cite{Gross:2015bea}. Furthermore, to avoid that the Higgs developing a super-Hubble mass during inflation, one needs {\small$g^2 \leq 10^{-4}g_*^2$}. We will consider this second inequality as valid from now on, providing in this way an operational definition of what we mean by the Higgs being sufficiently weakly coupled to the inflationary sector. 

The concrete particle physics realization of inflation has eluded any clear identification so far. The inflationary dynamics is normally described in terms of a scalar field, the inflaton, a singlet under the SM, and with a vacuum-like energy density. For simplicity, we will describe inflation as a {\it de Sitter} background with Hubble rate {\small$H_{*} \gg M_{\rm EW}$}, where {\small$M_{\rm EW} \sim \mathcal{O}(10^2)$} GeV is the electroweak (EW) scale. 
The current upper bound of the inflationary Hubble rate is~\cite{Ade:2015xua} 
\begin{eqnarray}
H_* \leq H_*^{\rm (max)} \simeq 8.5\times10^{13} {\rm GeV}\,,
\end{eqnarray}
so, in principle, there is 'plenty of room' to fulfill the demand {\small$M_{\rm EW} \ll H_* \leq H_*^{\rm (max)}$}. 

In the unitary gauge, the Higgs doublet can be written as {\small$\Phi = \varphi/\sqrt{2}$}, with {\small$\varphi$} a real degree of freedom with renormalized potential at large field values ({\small$\varphi \gg M_{\rm EW}$})
\begin{eqnarray}\label{eq:higgs-potential}
V = {1\over 4}\lambda(\varphi)\varphi^4\,,
\end{eqnarray}
where the effective self-coupling {\small$\lambda(\varphi)$}, encapsulates the radiative corrections to the Higgs potential~\cite{Casas:1994qy,Casas:1996aq}.

Under the circumstance that the Higgs is not responsible for Inflation, and is decoupled from (or weakly coupled to) the inflationary sector, we can consider two possibilities:

\begin{enumerate}[(i)]
\item {\it Higgs minimally coupled to gravity\,--}. In this case, the Higgs plays no dynamical role during inflation. It behaves as a light spectator field, independently of its initial amplitude~\cite{DeSimone:2012qr,Enqvist:2013kaa}. The Higgs performs a random walk at superhorizon scales during inflation, reaching within few e-folds an equilibrium distribution with variance~\cite{Starobinsky:1994bd} 
\begin{eqnarray}\label{eq:VarCaseI}
~~~~~~~\langle \varphi^2 \rangle \simeq 0.13{H_{*}^2\over\sqrt{\lambda}}\,.
\end{eqnarray}
The Higgs forms this way a condensate with a large VEV during inflation. A typical amplitude of the Higgs condensate is then {\small$\varphi_{\rm rms} \sim H_{*}$}, (almost) independently of the Higgs self-coupling for reasonable values of {\small$\lambda$}. The scale over which the Higgs condensate amplitude fluctuates, i.e., the correlation length of the Higgs condensate, is exponentially larger than the Hubble radius~\cite{Starobinsky:1994bd}, {\small$l_* \sim \rm{exp} (3.8/\sqrt{ \lambda}) H_*^{-1} \gg H_*^{-1}$}. The Higgs condensate is then homogeneous within cosmological scales. 

\item {\it Higgs non-minimally coupled to gravity\,--}. An interaction {\small$\xi \Phi^\dag\Phi R$}, with {\small$R$} the {\small$Ricci$} scalar, is required by the renormalisation of the SM in curved space. If the value of {\small$\xi$} at the inflationary scale lies below {\small$\xi \lesssim 0.1$}, the Higgs is sufficiently light during inflation, and hence we recover scenario {\it i)}. If on the contrary, {\small$\xi > 0.1$}, the Higgs becomes too heavy during inflation. It develops a super-Hubble mass and, consequently, it is not excited as a condensate. However, the sudden drop of the curvature {\small$R$}, during the transition from the end of inflation to a standard power-law post-inflationary regime, induces a non-adiabatic change in the effective Higgs mass {\small$ m_\varphi^2 = \xi R$}. This translates into a significant excitation of the Higgs modes at the Hubble scale {\small$k \sim a_*H_*$}. Following the recent analysis of~\cite{Herranen:2015ima}, the Higgs excitation acquires a large variance\footnote{Denoting the post-inflationary equation of state as {\small$w$}, in~\cite{Herranen:2015ima} it is found that {\small$\langle \varphi^2 \rangle\simeq \mathcal{O}(10^{-2})H_{*}^2/\sqrt{\xi}$} for {\small$w = 1/3$}, or {\small$\langle \varphi^2 \rangle \simeq \mathcal{O}(1)H_{*}^2\sqrt{\xi}$} for {\small$w \neq 1/3$}.} depending on the post-inflationary equation of state {\small$w$}. 
Furthermore, if after inflation the inflaton oscillates around the minimum of its potential, the periodic time-dependent behavior of {\small$R$} excites the Higgs modes through parametric resonance. The Higgs amplitude cannot in any case exceed~\cite{Herranen:2015ima} 
\begin{eqnarray}\label{eq:VarCaseII}
~~~~~~~\langle \varphi^2 \rangle \lesssim \mathcal{O}(0.1){H_{*}^2\over\lambda\sqrt{\xi}}\,,
\end{eqnarray}
as the Higgs self-interactions prevents any further growth above this value. In summary, in the presence of a large non-minimal coupling, the Higgs forms a condensate immediately after inflation, with a large amplitude bounded as {\small$\varphi_{\rm rms} \lesssim H_{*}/\lambda^{1/2}\xi^{1/4}$}. The coherent scale is, however, only of the size of the Hubble radius, {\small$l_* \lesssim H_*^{-1}$}, as the fluctuations at different horizon patches are uncorrelated.

\end{enumerate}

The running of the Higgs self-coupling {\small$\lambda(\mu)$} has been computed up to three loops~\cite{Degrassi:2012ry,Bezrukov:2012sa}. The self-coupling decreases with energy ({\small$d \lambda / d \mu < 0$}), becoming negative at a given critical scale {\small$\mu_c$}. Due to this, the Higgs potential Eq.~(\ref{eq:higgs-potential}) possesses a maximum (a barrier) at a scale {\small$\mu_+ \lesssim \mu_c$}, crosses zero at {\small$\mu_c$}, and it (possibly) develops  a negative minimum at higher energies. These scales are very sensitive to the Higgs mass {\small$m_H$}, the strong coupling constant {\small$\alpha_s$}, and the top quark mass {\small$m_t$}. For the SM central values for {\small$\alpha_s$} and {\small$m_H$}, as well as the world average top quark mass {\small$m_t = 173.34 \rm{GeV}$} \cite{ATLAS:2014wva}, one finds {\small$\mu_+ \approx 7 \times 10^{9} \rm{GeV}$} and {\small$\mu_c \approx 10^{10} \rm{GeV}$}. However, considering a value of {\small$m_t$} two/three sigma below its central value, we put the critical scales at {\small$\mu_+,\mu_0 \geq 5 \times 10^{16} \rm{GeV} \gg H_*^{\rm (max)}$}, which is one way of ensuring the stability. Another way is to consider Higgs portals to scalar fields, changing the running of {\small$\lambda$} so that it always remains positive~\cite{Ballesteros:2015iua}. In this work we assume that the Higgs potential is stable up to inflationary energies, so we consider that {\small$\lambda$} never becomes negative. The higher the energy scale of inflation, the smaller is $\lambda$, with reasonable values only within the interval {\small$10^{-2} \lesssim \lambda < 10^{-5}$}~\cite{Figueroa:2015hda} ({\small$\lambda \sim 10^{-5}$} being only marginally valid). 

\section{Post-inflationary dynamics of the SM Higgs and its decay products}
\label{sec:PostInfDynamics}

As we have just discussed, either during inflation [case $i)$] or just towards its end [case $ii)$], the Higgs is excited in the form of a condensate with a large VEV. As a consequence, the Higgs condensate starts oscillating around the minimum of its potential, soon after the end of inflation. Each time the Higgs crosses zero, all particles coupled to the Higgs, i.e.~the electroweak gauge bosons and the charged fermions of the SM, are created in non-perturbative bursts~\cite{Enqvist:2013kaa,Figueroa:2014aya,KariEtAlNonAbelian14,Figueroa:2015hda,Enqvist:2015nrw}. In the case of gauge bosons this phenomenon is called parametric resonance, and it is very similar to the decay process of an inflaton in preheating with a quartic potential~\cite{Greene:1997fu}. The main difference with respect to preheating is that, in the present scenario, the Higgs does not dominate the energy budget of the Universe. On the contrary, given the typical Higgs condensate amplitudes {\small$\varphi_{\rm rms}$} [Eq.~(\ref{eq:VarCaseI}) in case $i)$, or Eq.~(\ref{eq:VarCaseII}) in case $ii)$], the Higgs energy density is always much smaller than the energy density of the inflationary sector {\small$\rho_{*} \equiv 3 m_p^2 H_*^2$}, 
\begin{eqnarray}\label{eq:InitialRatioEnergies}
{{1\over4}\lambda \langle\varphi^4\rangle \over \rho_{*}} \lesssim \delta\times\mathcal{O}(10^{-12})\,\left({H_*\over H_*^{\rm (max)}}\right)^2~\ll~ 1\,,
\end{eqnarray}
with {\small$\delta = 1$} [case $i)$] or {\small$\delta \equiv {1/{\lambda\xi}}$} [case $ii)$].

The decay of the Higgs and the dynamics of its energetically dominant decay products have been recently studied under these circumstances, with the help of classical lattice simulations~\cite{Figueroa:2015hda,Enqvist:2015nrw}. The lattice approach takes into account the non-linearities of the system beyond the analytical approach (based on the initially valid linear regime). Strictly speaking, however, the analysis presented in~\cite{Figueroa:2015hda,Enqvist:2015nrw} only describes the post-inflationary dynamics of the system in the spectator field case $i)$, when the Higgs is excited as a condensate during inflation. 

The post-inflationary dynamics in the case $ii)$, when a non-minimal coupling to gravity {\small$\xi\varphi^2R$} is present, can however be expected to be (qualitatively) similar to that of case $i)$, at least for moderate values of $\xi$. The reason for this is that the coupling to the scalar curvature {\small$R$}, introduces an additive term in the Higgs oscillation frequency, proportional to the square of the Hubble rate, {\small$R \propto H^2$}. As {\small$H^2 \sim 1/t^2$}, the new term decays rapidly in time. Therefore, the Higgs oscillations in the case $ii)$ are expected to be initially modulated by the non-minimal coupling to gravity, but tending rapidly to the oscillations of case $i)$ (characterized by an oscillation frequency $\propto \lambda\varphi^2$, given by the Higgs self-interactions).

In what follows, we will describe the details of the post-inflationary dynamics, considering only the scenario $i)$ with $\xi$ set to zero. This should capture equally well the dynamics when $0 < \xi \lesssim 0.1$. For the scenario $ii)$ with a non-minimal coupling $\xi > 0.1$, we expect the dynamics to tend rapidly to the case of $\xi < 0.1$, unless $\xi$ is extremely large. When $\xi \gg 1$, a new analysis beyond our current study should be performed. However, the larger the $\xi$, the smaller the GW production is expected to be\footnote{The larger the $\xi$, the faster the energy transfer from the Higgs into its decay products, as the particle production (due to the Higgs oscillations) is expected to be modulated by a larger oscillation frequency. As we will explain later on in Section~\ref{sec:GW}, the faster the decay of the Higgs proceeds, the higher the frequency of the GW produced during the process. However, the amount of GW produced is mostly determined by the energy stored initially in the Higgs condensate, and not by the rapidity of its decay. In the scenario $ii)$, the larger the $\xi$, the smaller the initial energy stored in the Higgs~\cite{Herranen:2015ima}. Therefore, the larger the $\xi$, the larger the frequency of the GW background, but the smaller the amplitude of the background. We expect therefore the GW production in scenario $i)$ with $\xi = 0$ to be the largest possible one.}. Therefore, we only focus from now on in the details of scenario $i)$ with $\xi = 0$, which maximizes the GW production. This should still capture qualitatively well the dynamics of the system, even in the presence of a moderate non-minimal coupling to gravity, as long as $\xi$ is not extremely large.

\subsection{Abelian model of the electroweak interactions}

In order to describe the production of gauge bosons from the Higgs decay after inflation, we will follow the approach presented in~\cite{Figueroa:2015hda}. We will model the electroweak sector of the SM with an Abelian-Higgs set-up, ignoring the non-linearities arising from the full non-Abelian structure of the electroweak interactions. Non-abelian corrections over the Abelian dynamics are indeed suppressed as $\propto 1/\sqrt{q}$~\cite{KariEtAlNonAbelian14,Figueroa:2015hda,Enqvist:2015nrw}, where {\small$q \equiv e^2/\lambda$} is the resonance parameter of the gauge field(s), and {\small$e^2$} represents the Abelian coupling (mimicking either of the {\small$W^\pm$} or {\small$Z$} gauge couplings). The Abelian approximation to the full Higgs-gauge electroweak interactions works better the larger is the resonance parameter $q$. For GW production through the Higgs decay products, we are interested in inflationary scenarios with the highest possible energy scales, since this enhances the GW produced, see Section~\ref{sec:GW}. If the inflationary Hubble rate {\small$H_*$} is sufficiently high, say of the order of (though somewhat smaller than) its current upper bound {\small$H_* \lesssim H_* ^{\rm (max)} \sim 10^{14}$} GeV, the value of {\small$\lambda$} runs into small values. This implies that the resonance parameters are, in fact, rather large, {\small$q \equiv {e^2\over \lambda} \gg 1$}. Therefore, taking the Abelian approximation in this regime is well justified\footnote{Lattice simulations considering the non-Abelian structure of the
electroweak interactions in a different post-inflationary context have been done in~\cite{GarciaBellido:2003wd}. In the scenario we study in this paper, simulations considering the non-Abelian structure were presented in~\cite{Enqvist:2015nrw}. These simulations show, however, that even for the lowest possible resonance parameter(s) {$q \lesssim 10$} (for which the non-Abelian effects are maximized), the time scales of the problem are modified with respect the Abelian approximation only by a factor of {$\sim 2$}.}, and hence in the present work we will consider the Abelian modeling from now on. 

Within the Abelian-Higgs modeling, the interactions of the Higgs with a gauge boson species $A_\mu$ can be described by the action {\small$S = \int \mathcal{L}\, d^4x$}, with
\be
- \mathcal{L} =   (D_{\mu} \Phi)^* (D^{\mu} \Phi ) + \frac{1}{4 e^2} F_{\mu \nu} F^{\mu \nu} + \lambda (\Phi^* \Phi )^2\,. \label{action}
\ee
Here {\small$D_{\mu} \equiv \partial_{\mu} - i A_{\mu}$} is the covariant gauge derivative, {\small$F_{\mu \nu} \equiv \partial_{\mu} A_{\nu} - \partial_{\nu} A_{\mu}$} is the (Abelian) field strength, and {\small$e$} is the strength of the Abelian coupling. In the Abelian modeling the Higgs needs to be considered as a complex scalar field {\small$\Phi \equiv \frac{1}{\sqrt{2}} \varphi = \frac{1}{\sqrt{2}} ( \varphi_1 + i \varphi_2 )$}, with {\small$\varphi_i \in \mathfrak{R}$}. As we are dealing with a gauge theory, we have a gauge freedom to choose the field components. This allows us to set the gauge {\small$A_0=0$} from now on. By varying the action, the equations of motion can be derived as
\bea \ddot{\Phi} - D_i D_i \Phi + 2{\dot a\over a} \dot{\Phi}   &=& - 2 \lambda a^2 (t) |\Phi|^2\Phi \ , \label{eom1} \\
\ddot{A}_j + \partial_j \partial_i A_i - \partial_i \partial_i A_j &=& 2 e^2 a^2 (t) \mathfrak{Im} [ \Phi^* D_j \Phi ] \ ,\label{eomb}   \\ 
\partial_i \dot{A}_i &=& 2 e^2 a^2(t)  \mathfrak{Im}[\Phi^* \Phi] \ . \label{eomc} \eea
Eqs.~(\ref{eom1}) and (\ref{eomb}) of the system describe the dynamics of the Higgs and the gauge boson, while Eq.~(\ref{eomc}) is the Gauss law, representing a constraint that must be obeyed at all times. The gauge-invariant electric and magnetic fields are {\small$E_i \equiv F^{0 i}$} and {\small$B_i = \frac{1}{2} \epsilon_{i j k} F^{j k}$}.
 
We model the scale factor as 
\be a(t) = a_* \left( 1 + \frac{1}{2}(1+3\omega)a_* H_* (t - t_*) \right)^\frac{2}{1 + 3 \omega} \ , \label{eq:exprate} \ee
where {\small$w$} is the equation of the state of the universe after inflation. The value of $w$ depends on the inflationary sector, which is not specified here, so we will consider various values representing different expansion rates. For example, for matter-domination (MD), radiation-domination (RD) or kination-domination (KD), the equation of state is {\small$w= 0, {1\over3}, 1$}, respectively. From now on we fix {\small$a_* = a(t_*) \equiv 1$}.

It is convenient to define dimensionless spacetime variables {\small$z^{\mu} = (z^0, z^i)$} as
\be z \equiv z^0 = H_* t \,,~~~ z^i = H_* x^i \label{def-st2} \ . \ee
It is also convenient to define dimensionless Higgs and gauge field amplitudes as
\be h \equiv  \frac{a(z)}{a_*} \frac{\varphi}{\varphi_*} \ ,  \hspace{1cm} V_{i} \equiv \frac{1}{H_*} A_{i}  \ , \label{defh} \ee
({\small$i=1,2,3$}) with {\small$h \equiv h_1 + i h_2$}, where {\small$\varphi_* \equiv |\varphi (t_*)|$} is the initial modulus of the Higgs field at the end of inflation. To distinguish between variables, we use a dot or a prime to denote differentiation with respect to conformal or natural variables, {\small$~\dot{} \equiv d/dt$} or {\small$, ~' \equiv d/dz$}. From now on, all spatial derivatives will also be taken with respect the new variables, unless otherwise stated. We also define a dimensionless covariant derivative as {\small$\mathcal{D}_i \equiv \frac{ \partial}{\partial z_i} - i V_i$}. With these changes, Eqs.~(\ref{eom1})-(\ref{eomc}) can be written as
\bea 
h''  - \mathcal{D}_{i} \mathcal{D}_{i} h + \beta^2 |h|^2 h  &=& h \frac{a''}{a} \ ,   \label{eomb1}\\
V_j'' + \partial_{j} \partial_{i} V_i - \partial_{i} \partial_{i} V_j &=&  q \beta^2  \mathfrak{Im} [ h^* \mathcal{D}_{i} h ] \ ,  \label{eomb3}\\
\partial_i V_{i}' &=&  q \beta^2  \mathfrak{Im} [ h^* h' ] \ , \label{eomb4}  
\eea
where {\small$q \equiv \frac{e^2}{\lambda}$} is the resonance parameter, and we have defined the parameter
\be \beta \equiv \frac{\sqrt{\lambda} \varphi_*}{H_*} \label{eq:beta}\ ,\ee
characterizing the initial Higgs amplitude.

Let us denote by {\small$W_{\mu}^+$}, {\small$W_{\mu}^-$} and {\small$Z_{\mu}$}, the Abelian version of the SM electroweak {\small$W, Z$} gauge bosons. Let us also denote by {\small$g_W$} and {\small$g_Z$} the SM gauge couplings of the Higgs to such bosons. In order to mimic correctly the real Higgs-gauge interactions, we need to identify $e^2 = g^2/4$ in Eqs.~(\ref{eom1})-(\ref{eomc}), with {\small$g^2 = g_W^2$} for {\small$W$} bosons and {\small$g^2 = g_Z^2$} for {\small$Z$} bosons. This translates into the following resonance parameters in Eqs.~(\ref{eomb1})-(\ref{eomb4}),
\begin{eqnarray}
q_Z \equiv \frac{g_Z^2}{4 \lambda}\,,~~~ q_W \equiv \frac{g_W^2}{4 \lambda}\, . 
\end{eqnarray}

In principle, for each of the three gauge bosons {\small$W^\pm$} and {\small$Z$}, there should be an equation of motion like Eq.~(\ref{eomb}) [or equivalently Eq.~(\ref{eomb3})], and a Gauss constraint like Eq.~(\ref{eomc}) [or equivalently Eq.~(\ref{eomb4})]. However, we demonstrated in~\cite{Figueroa:2015hda} that this system of three gauge fields can be mapped identically into another system with a single gauge boson, defined as
\be \label{eq:superboson}  S_{\mu} \equiv W_{\mu}^+ + W_{\mu}^- + Z_{\mu}\,, \ee
with associated resonance parameter (gauge coupling)
\be \label{eq:superCoupling} q_s \equiv q_Z + 2 q_W = \frac{g_Z^2 + 2 g_W^2}{4 \lambda} \ . \ee
See Section V.B.~in~\cite{Figueroa:2015hda} for details. At any moment one can recover the original fields by simply taking {\small$W_{\mu}^- = W_{\mu}^+ = (q_W / q_s) S_{\mu}$} and {\small$Z_{\mu} = (q_Z / q_s) S_{\mu}$}. At very high energies, the running of the SM gauge couplings show that {\small$g_Z^2 \approx 2 g_W^2$}, so {\small$q_s \approx 2 q_Z \approx 4 q_W$}, and hence {\small$W_{\mu}^- = W_{\mu}^+ \approx Z_{\mu} / 2$}.

The energy density of the Higgs + gauge fields is
\be \rho_{\rm s} (z) =  \frac{V_*}{a^4 (z)} E_s (z)\,,~~~~V_* \equiv {\lambda\over 4} |\varphi_*|^4\,, \label{rhoz_gauge}\ee
where $V_*$ is the value of the Higgs potential at the end of inflation, and the function $E_s (z)$ is formed by the sum of the following energetic contributions:
\be E_s (z) = E_{\rm K} + E_{\rm V} + E_{\rm E} + E_{\rm M} + E_{\rm GD}\ . \label{rhot-gauge} \ee
Here, $E_{\rm K}$ and $E_{\rm V}$ are the kinetic and potential energies of the Higgs field, and $E_{\rm E}$ and $E_{\rm M}$ are the electric and magnetic energy densities of the super gauge-boson $S_\mu$ of Eq.~(\ref{eq:superboson}) (hence containing the contribution from all the gauge bosons $W^\pm, Z$),
\bea E_{\rm K}^{\varphi} &\equiv& \frac{a^4}{V_*} \frac{\sum_i \dot{\varphi}_i^2}{2 a^2} = \frac{2}{\beta^2} \sum_i^{2}  \left( h'_i - h_i \frac{a'}{a} \right)^2 \label{eq:kinetic-energy}  \ , \\
E_{\rm V} &\equiv& \frac{a^4}{V_*} \frac{\lambda ( \varphi_1^2 + \varphi_2^2 )^2}{4}= (h_1^2 + h_2^2)^2 \ ,\\ 
E_{\rm E} &\equiv& \frac{a^4}{V_*} \frac{1}{2 e^2 a^4 } \sum_i E_i^2 = \frac{2 }{q \beta^4} \sum_i \mathcal{E}_i^2 \label{eq:electric-energy} \ , \\
E_{\rm M} &\equiv& \frac{a^4}{V_*} \frac{1}{2 e^2 a^4 } \sum_i B_i^2 =  \frac{2}{q \beta^4} \sum_i \mathcal{B}_i^2 \label{eq:magnetic-energy} \ ,\eea
where we have defined dimensionless electric and magnetic fields as {\small$\mathcal{E}_i = E_i /H_*^2$} and {\small$\mathcal{B}_i = B_i / H_*^2$}. The last contribution {\small$E_{\rm GD}$} is a gauge-invariant term formed by the product of covariant derivatives of the Higgs field,  
\bea  E_{\rm GD} &\equiv& \frac{a^4}{V_*} \frac{1}{2 a^2} \sum_i \mathfrak{Re}  [ (D_i (\varphi_1 + i \varphi_2))^* D_i (\varphi_1 + i \varphi_2) ] \nonumber \\
&=& \frac{2}{\beta^2} \sum_i \mathfrak{Re}  [ (\mathcal{D}_i (h_1 + i h_2))^* \mathcal{D}_i (h_1 + i h_2) ] \ ,
\label{eq:GradientEnergy} \eea
hence representing the energy stored in the spatial Higgs gradients and the Higgs-gauge interactions.

\begin{figure}[t]
    \begin{center}
        \includegraphics[width=8cm]{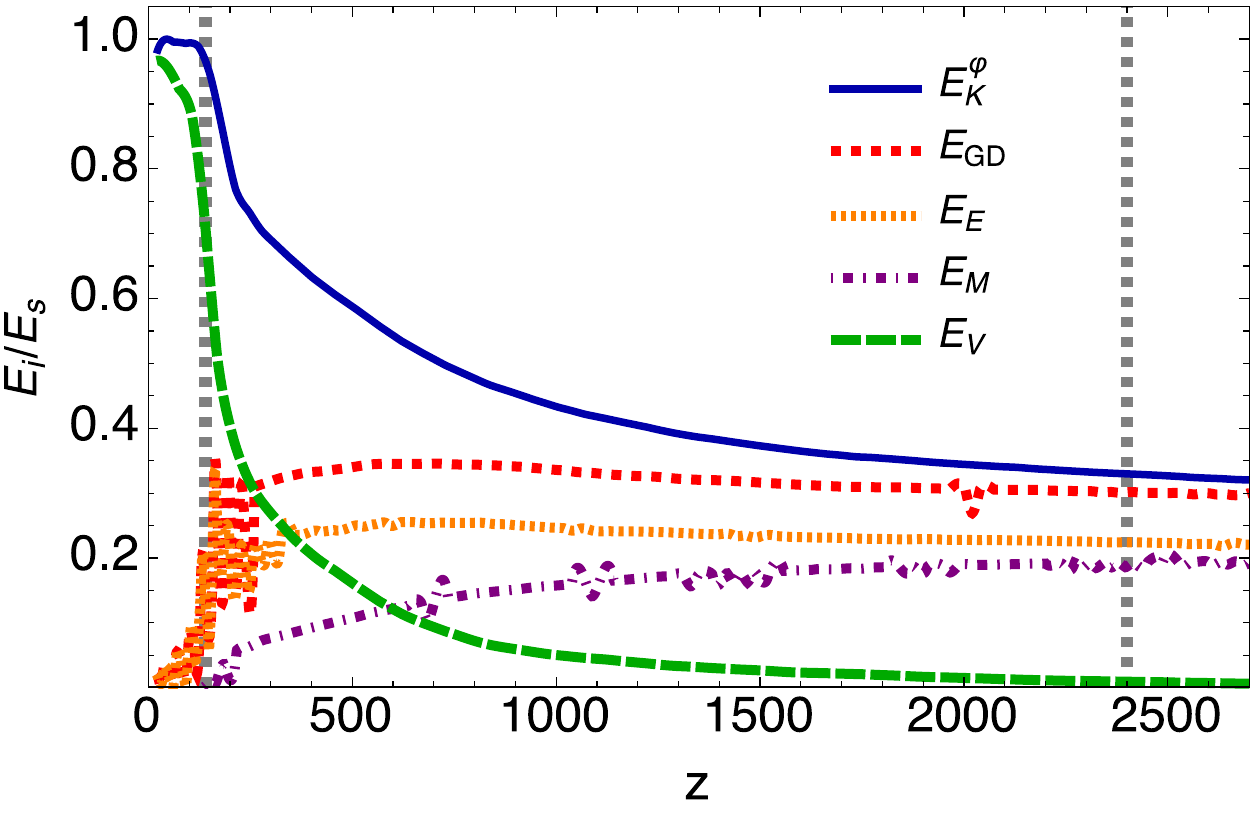}
    \end{center}
    \caption{Plot, as a function of time, of the different contributions {\small$E_i / E_s$} [see Eq.~(\ref{rhot-gauge})] to the total energy of the system, obtained from a lattice simulation for {\small$q_s=79$}, a RD post-inflationary expansion, and $\beta=0.01$. All functions are oscillating, but we take the envelope of the corresponding oscillations for clarity. The first dashed vertical line indicates the time {\small$z_i (q)$}, whereas the second dashed vertical line indicates the time {\small$z_e(q)$}.} 
    \label{fig:Energy-Components}
\end{figure}

We have plotted in Fig.~\ref{fig:Energy-Components} the different energy contributions as a function of time for the resonance parameter {\small$q_{\rm s} = 79$}. We have normalized each energy component as {\small$E_i / E_s$}, and we have removed their oscillations, hence showing only the corresponding envelope functions. We see that initially the dominant contributions come from the kinetic and potential energies of the Higgs field. This corresponds to the oscillations of the Higgs condensate around the minimum of its potential, before the gauge fields backreact onto the Higgs. Meanwhile, the other components of the energy ({\small$E_{\rm E}$}, {\small$E_{\rm M}$} and {\small$E_{\rm GD}$})  grow very fast, due to the energy transfer -- via parametric resonance -- from the Higgs to the gauge fields. We can understand the time evolution of these energies in light of the context of the next subsection. 

\subsection{Summary of the post-inflationary dynamics of the Higgs and its decay products}

The lattice simulations presented in~\cite{Figueroa:2015hda} represent a precise calculation of the dynamics of the SM after inflation, fully incorporating the nonlinear and non-perturbative effects of the SM gauge interactions between the Higgs and the {\small$W^\pm$} and {\small$Z$} gauge bosons, though neglecting the truly non-Abelian structure. The advantage of a lattice approach with respect an analytical one is that the first includes the effect of the non-linearities of the system into the Higgs decay dynamics, which become relevant very soon after the end of inflation. Here we just briefly summarize the results obtained in~\cite{Figueroa:2015hda}.

The production of the gauge bosons from the post-inflationary Higgs decay is controlled by three parameters: the amplitude of the Higgs condensate {\small$\varphi_*$} at the end of inflation, the equation of state {\small$w$} of the Universe in the period following immediately after inflation, and the resonance parameter {\small$q \equiv g^2/4\lambda$} of the gauge fields, with {\small$g^2$} equal to either of the {\small$W^\pm, Z$} gauge couplings\footnote{The production of fermions is equally controlled by the same parameters, {\small$\varphi_*, w, q$}, but substituting the gauge coupling {\small$g^2$} in the definition of {\small$q$} by the Yukawa coupling {\small$y_i^2$}, with the {\small$i$}-index referring to the fermion species~\cite{Figueroa:2014aya}.}. Taking the end of inflation as the initial time {\small$z_* = 0$}, the dynamics of the system at later times {\small$z \geq z_*$} can be characterized by three time scales:\vspace*{0.1cm} 

\begin{itemize}

\item {\small$z = z_{\rm osc}$}: This time signals the moment after inflation when the Higgs effective mass becomes larger than the Hubble rate. This 'forces' the amplitude of the Higgs condensate to start rolling down its potential. Previous to this moment, during {\small$z_* \leq z \leq z_{\rm osc}$}, the Higgs amplitude remains frozen in slow-roll\footnote{In the scenario $ii)$ with {\small$\xi > 0.1$}, the Higgs starts rolling down its potential immediately after inflation, since the effective mass of the Higgs (given by its non-minimal coupling to gravity) is larger than the Hubble already at $z_*$. Hence, {\small$z_{\rm osc} = z_{*} = 0$} in this case.} . Hence {\small$z = z_{\rm osc}$} signals the onset of the Higgs oscillations around the minimum of its potential. Every time the Higgs crosses around zero, particle creation via parametric effects occur. It is found that {\small$z_{\rm osc} \lesssim \mathcal{O}(10)$}, the exact value depending on the particular Higgs amplitude {\small$\varphi_*$} and post-inflationary equation of state {\small$w$}. Therefore, the Higgs oscillations start, in all cases, shortly after inflation ends. From then on, top quarks and gauge bosons are strongly created every time the Higgs crosses through zero. After few oscillations, the gauge boson production dominates, as the gauge fields develop parametric resonance, whereas the fermions are Pauli blocked~\cite{GarciaBellido:2001cb}. The excitation of the dominant species {\small$Z, W^\pm$} is very similar to the excitation of preheat fields coupled to the inflaton, as preheating is due to oscillations of an inflaton with quartic potential.

\item {\small$z = z_i$}: This second time scale signals the moment when the produced gauge bosons have accumulated sufficient energy such that they start backreacting onto the Higgs condensate, starting to affect the latter severely. At $z \geq z_i$ there is a sharp decrease of the amplitude and energy density of the Higgs condensate. This scale depends strongly on the resonance structure of the gauge field dynamics, characterized by the particular value of the resonance parameter {\small$q_s \equiv (g_Z^2 + 2 g_W^2)/ (4 \lambda)$}. For the larger resonance parameters (due to small values of $\lambda$), {\small$z_i$} tends to be shorter, while weaker-resonance parameters cases have larger {\small$z_i$}'s. In practice, {\small$z_i \lesssim z_{\rm osc} + (\mathcal{O}(0.1)-\mathcal{O}(10^3))$}. 

\item {\small$z = z_e$}: This third and final time scale signals the end of the transfer of energy from the Higgs into the SM species, as well as the onset of a stationary regime. The time {\small$z_e$} depends on the resonance parameter as a power law {\small$\propto q_s^\alpha$}, where a phenomenological fit to the simulations shows {\small$\alpha \sim 0.42$}~\cite{Figueroa:2015hda}. Therefore, the greater the {\small$q_s$}, the longer the time {\small$z_e$}. At {\small$z \geq z_e$}, the energies of the different components of the system have established an equipartition regime, with fixed relative ratios of the energies independent of {\small$q_s$}. For reasonable values of $\lambda$, $q_s$ ranges between {\small$\sim \mathcal{O}(10)$} and {\small$\sim \mathcal{O}(10^3)$}, so {\small$z_e$} ranges between {\small$\sim 10^3$} and {\small$\sim 10^4$}. 

\end{itemize}

\section{Gravitational wave production}
\label{sec:GW}

Gravitational waves (GW) are tensor perturbations which propagate following the equation of motion~\cite{Maggiore2000}
\be \ddot{h}_{i j} + 2{\dot a\over a} \dot{h}_{ij} - \partial_k\partial_k h_{i j} = {2\over m_p^2}\Pi_{i j}^{\rm TT}\,, \label{eq:gw}\ee
where the source of GW, {\small$\Pi_{i j}^{\rm TT}$}, is the {\it transverse}-{\it traceless} (TT) part of the anisotropic stress tensor {\small$\Pi_{ij}$}. In our case, in the presence of both scalar and vector fields, the source is effectively given by~\cite{DaniPhD}
\bea \Pi_{i j}^{\rm TT} &=& \left\lbrace 2 \mathfrak{Re} [ (D_i \varphi)^* (D_j \varphi ) ] + \frac{1}{e^2 a^2} \eta^{\alpha \beta} F_{i \alpha} F_{j \beta} \right\rbrace^{\rm TT} \nonumber\\
&=& \left\lbrace 2 \mathfrak{Re} [ (D_i \varphi)^* (D_j \varphi ) ] - \frac{1}{e^2 a^2} (E_{i}E_{j}+B_iB_j) \right\rbrace^{\rm TT} \hspace*{-3mm},\nonumber \\ \label{eq:Projector}
\eea
where {\small$\lbrace ... \rbrace^{\rm TT}$} represents the TT part of the quantity inside the brackets, {\small$D_\mu$} is the gauge covariant derivative in Eq.~(\ref{action}), {\small$E_i \equiv F^{0 i}$} and {\small$B_i = \frac{1}{2} \epsilon_{i j k} F^{j k}$} are the electric and magnetic fields formed from the superfield $S_\mu$ in Eq.~(\ref{eq:superboson}), and we have discarded a term {\small$\propto (|{E}|^2 + |{B}|^2 ) \delta_{i j}$} because it is actually a pure trace term. Note that here the charge must be identified with that of $S_\mu$, so that {\small$e^2 \equiv (2g_W^2+g_Z^2)/4$}.

It is convenient to redefine the tensor mode amplitude through a conformal redefinition like (recall that initially we take $a_* = 1$)
\be h_{ij} \equiv {\bar{h}_{ij}\over a}\,,\ee
so that Eq.~(\ref{eq:gw}) can be written in terms of the dimensionless variables Eqs.~(\ref{def-st2}),(\ref{defh}) as
\bea \bar{h}_{ij}'' - \left(\partial_k\partial_k + \frac{a''}{a}\right)\bar{h}_{ij} = \frac{2}{a} {1\over\lambda}\left({H_*\over m_p}\right)^2\mathcal{P}_{ij}^{\rm TT} \ , \label{eq:gw-eom}
\eea
with
\vspace*{-2mm}
\begin{gather}
\mathcal{P}_{ij} = \mathcal{P}_{ij}^{[h]} + \mathcal{P}_{ij}^{[g]}\\
\mathcal{P}_{ij}^{[h]} \equiv \beta^2 \mathfrak{Re}[ (D_i h)^* (D_j h) ]\,,~~\mathcal{P}_{ij}^{[g]} \equiv - \frac{1}{q_s} ( \mathcal{E}_i \mathcal{E}_j + \mathcal{B}_i \mathcal{B}_j )\label{eq:GWsource}\,.
\end{gather}
We recall that $\mathcal{E}_i \equiv E_i/H_*^2, \mathcal{B}_i \equiv B_i/H_*^2$, and {\small$q_s = (2g_W^2+g_Z^2)/4$} is the total resonance parameter Eq.~(\ref{eq:superCoupling}). It is clear from Eq.~(\ref{eq:GWsource}) that both the Higgs field and the gauge bosons will contribute as a source of GW.

The spectrum of the GW energy density contained within a volume $V$, and normalized to the total energy density $\rho_{\rm tot}$ of the Universe (at the time of GW production), can be written in the continuum as
\bea \Omega_{_{\rm GW}}(k,z) &\equiv &
\frac{1}{\rho_{\rm tot}}\frac{d \rho_{_{\rm GW}}}{d \log{k}}(k,z)\\
& = & \frac{1}{8\pi^2 a^2}\frac{m_p^2\,k^3}{\rho_{\rm tot}V} \left\langle \dot{h}^{*}_{ij}(k,z)\dot{h}_{ij}(k,z) \right\rangle_{\hspace*{-0.5mm}4\pi}\nonumber
\eea
where {\small$\left\langle ... \right\rangle_{\hspace*{-0.2mm}4\pi} \equiv {1\over 4\pi}\int\hspace*{-0.5mm} d\Omega_k\,...$}, with {\small$d \Omega_k$} a solid angle differential in {\bf k}-space. 

In light of the parameters factorized out in the source term of Eq.~(\ref{eq:gw-eom}), it is convenient to define a new variable 
\bea
\bar{h}_{ij}(k,z) \equiv {2\over\lambda}\left({H_*\over m_p}\right)^2 w_{ij}(k,z)\,,
\label{eq:NewHij}
\eea

It is then useful to express {\small$\Omega_{_{\rm GW}}(k,z)$} in terms of the natural variables of the problem {\small$z_\mu$} and {\small$w_{ij}$}. We can factorize out this way the dependence with the Hubble scale {\small$H_*$} and the background expansion rate as
\bea
\Omega_{_{\rm GW}}(k, z) \equiv \delta_{*}\,\epsilon_w(a)\,\Theta_{_{\rm GW}}(k,z)\,, \label{eq:TotGW}
\eea
where

\bea
\delta_{*} \equiv \left(\frac{H_*}{m_p}\right)^4\,,~~~~~
\epsilon_w(a) \equiv  \left(a\over a_*\right)^{3w-1}\label{eq:GWfactorsOUT}
\eea
and
\bea
\Theta_{_{\rm GW}} (k,z) \equiv \frac{k^3}{6\pi^2  \lambda^2} \frac{1}{V} {\left\langle (w_{ij}' - \mathcal{H} w_{ij})(w_{ij}' - \mathcal{H} w_{ij})\right\rangle_{\hspace*{-0.5mm}4\pi} }\,.\label{eq:theta-defn}\nonumber\\ 
\eea

In order to derive Eqs.~(\ref{eq:TotGW})-(\ref{eq:theta-defn}), we have used that the total energy density of the Universe can be expressed as {\small$\rho_{\rm tot} = 3m_p^2H_*^2a^{-3(1+w)}$}, with $w$ the post-inflationary equation of state. We recall as well that {\small$\mathcal{H} \equiv a'/a$}. The factorization {\small$\Omega_{_{\rm GW}} = \delta_{*} \epsilon_w \Theta_{_{\rm GW}}$} in Eq.~(\ref{eq:TotGW}) is indeed very convenient: the dependence on {\small$\lbrace q_s,\beta,w\rbrace$} of {\small$\Theta_{_{\rm GW}}(k,z)$}, comes only from the effect of these parameters on the dynamics of the $eom$ of the Higgs + gauge fields system, Eqs.~(\ref{eomb1})-(\ref{eomb4}).

Note that the prefactor {\small$\delta_{*}$} in Eq.~(\ref{eq:TotGW}), implies a suppression of the GW (energy density) as {\small$\sim (H_* / m_p)^4 \lll 1$}. This effect is related to the fact that the typical initial amplitude of the Higgs condensate is {\small$\varphi_*^2 \sim \varphi_{\rm rms}^2 \sim H_*^2$}, which is then suppressed by the appearance of a Planck mass factor as {\small${1/m_p^2}$} in the $rhs$ of the GWs' Eq.~(\ref{eq:gw}). The scaling {\small$\propto \delta_{*}$} is ultimately responsible for the smallness of the GW background today, as we will emphasize later on in section~\ref{subsec:GWtoday}. Note that in standard preheating scenarios, say after chaotic inflation, the inflaton and preheat fields dominate the energy budget of the universe, and have typically much larger field amplitudes. Therefore, there is no such suppression in standard preheating via parametric resonance. The production of GW from subdominant field(s), like inflationary spectator fields as in our case, will however be always suppressed by the smallness of the fields amplitude {\small$\varphi \sim H_* \ll m_p$}. 

Depending on whether the post-inflationary equation of state is stiff, {\small$w > 1/3$}, or not, {\small$w \leq 1/3$}, the background energy density of the Universe will correspondingly decrease slower or faster than relativistic species, i.e.~{\small$d\log\rho_{\rm tot}/d\log a \propto -3(w+1)$} will be {\small$< -4$} for stiff backgrounds, or {\small$\geq -4$} for non-stiff backgrounds. The prefactor {\small$\epsilon_w = (a/a_*)^{3w-1}$} in Eq.~(\ref{eq:theta-defn}) will, therefore, either suppress the GW background as {\small$\propto \epsilon_w < 1$} for {\small$w < 1/3$} (e.g.~{\small$w = 0$} for MD), or enhance it as {\small$\propto \epsilon_w > 1$} for {\small$w > 1/3$} (e.g.~{\small$w = 1$} for KD). For {\small$w = 1/3$} the background energy density corresponds to a RD Universe, and hence {\small$\epsilon_{w} = 1$}, so that there is neither a suppression nor an enhancement. As we will discuss in section~\ref{subsec:GWtoday}, in a KD scenario with {\small$w = +1$}, the amplitude of the GW background will be maximally enhanced since {\small$\epsilon_w \gg 1$}. However, even in this case, the large suppression due to {\small$\delta_{*} \ll 1$} will still dominate over this enhancement, so that the overall modulation of the signal is {\small$\Omega_{_{GW}} \propto \delta_{*}\,\epsilon_{w} \sim (H_*/m_p)^2$}, which still represents a suppression, though a milder one.

In order to solve the $eom$ Eq.~(\ref{eq:gw-eom}) for the GW, we have followed the standard procedure first introduced in~\cite{GWsPRD2007FigueroaGarciaBellidoSastre}, solving a relativistic wave-like equation in real space sourced by the full {\small$\mathcal{P}_{ij}$}, with no TT projection,
\be u_{ij}''  - \left(\partial_k\partial_k + \frac{a''}{a}\right) u_{ij} = {1\over a}\mathcal{P}_{ij}\,. \label{eq:GW-nonTTeom}\ee
We can then recover {\small$w_{ij}$} at any moment, in Fourier space, through the relation
\begin{eqnarray}
w_{ij}(k,z) = u_{ij}^{\rm{TT}}(k,z) = \Lambda_{ij,lk}(\hat k)u_{lk}(k,z)\\
\Lambda_{ij,lk}(\hat k) = P_{il}P_{jk} - {1\over2}P_{ij}P_{lk}\,,~~P_{ij} = \delta_{ij} - \hat{k}_i\hat{k}_j \label{eq:TTprojector}
\eea
where {\small$\Lambda_{ij,lk}(\hat k)$} is a geometrical projector that filters out the TT degrees of freedom in Fourier space. Since {\small$\Lambda_{ij,pq}(\hat k)\Lambda_{pq,lm}(\hat k) = \Lambda_{ij,lm}(\hat k)$}, the argument inside the angular-average {\small$\langle ... \rangle_{\hspace*{-0.2mm}}$} in Eq.~(\ref{eq:theta-defn}) can be computed as

\begin{gather}
\left(w_{ij}'(k,z) - \mathcal{H} w_{ij}(k,z)\right)\left(w_{ij}'(k,z) - \mathcal{H} w_{ij}(k,z)\right) =\nonumber\\
\left(u_{ij}'(k,z) - \mathcal{H} u_{ij}(k,z)\right)\Lambda_{ij,lm}(\hat{k})\left(u_{lm}'(k,z) - \mathcal{H} u_{lm}(k,z)\right)
\end{gather}

We have studied the GW creation process in lattices of {\small$N = 256$} points per dimension. To solve the Higgs + gauge fields {\small$eom$} Eqs.~(\ref{eomb1}),(\ref{eomb3}), while verifying the constraint Eq.~(\ref{eomb4}) at every time, we have used the non-compact Abelian lattice formulation presented in Ref.~\cite{Figueroa:2015hda}. The exact details of the discretization formalism preserving gauge invariance are described in the Appendix A of~\cite{Figueroa:2015hda}, so we do not repeat them here. In the same way, a discussion of how we set the initial conditions of the different fields can be found in the Appendix B of the same paper. In all simulations we have ensured that the lattice resolution covers well the dynamical range of momenta excited in the process, for both the matter and the GW fields.

The discrete version of the GW $eom$ Eq.~(\ref{eq:GW-nonTTeom}), contrary to the matter fields $eom$ Eqs.~(\ref{eomb1})-(\ref{eomb4}), does not follow from a discretized action. Instead, we simply substituted the continuous derivatives {\small$\partial_\mu$} in Eq.~(\ref{eq:GW-nonTTeom}), with standard forward/backward lattice derivatives. In order to introduce a lattice version of the energy density spectrum of GW Eq.~(\ref{eq:theta-defn}), we followed the prescription introduced in \cite{Figueroa:GWformalism}. In our case, this translates into
\bea &&\Theta_{_{\rm GW}} ({\bf \tilde{n}}, z) = \frac{1}{6\pi^2 \lambda^2}{d{\tilde x}^3\,\kappa({\bf \tilde{n}})^3\over N^3}\\
&& ~~~~~~~~~~\times \left\langle (u_{ij}' - \mathcal{H} u_{ij})\Lambda^{\rm (L)}_{ij,lm}(u_{lm}' - \mathcal{H} u_{lm})\right\rangle_{\hspace*{-0.5mm}4\pi}\,, \label{eq:thetaGW}\nonumber\eea
where {\small$d{\tilde x} \equiv H_*dx$} is the dimensionless lattice spacing, {\small$\kappa({\bf \tilde{n}}) \equiv k({\bf \tilde{n}}) / H_*$} the dimensionless momenta, {\small$k({\bf \tilde{n}}) \equiv (2\pi/L)|{\bf \tilde{n}}|$} the momentum at the Fourier lattice site {\small${\bf \tilde{n}}$}, {\small$L$} the length of the lattice box, and {\small$w_{ij} \equiv w_{ij}({\bf \tilde{n}},z)$} the discrete Fourier transform of {\small$w_{ij}({\bf n},z)$}, with {\small${\bf n}$} labeling the lattice sites. Note that {\small$\Lambda^{\rm (L)}_{ij,lm}$} is a discretized version of the TT projector given in Eq.~(\ref{eq:TTprojector}), and multiple choices are possible. We have chosen a lattice projector based on forward derivatives, noticing that other choices did not change the GW spectra appreciably, see~\cite{Figueroa:GWformalism} for a thoughtful discussion on this point.

\section{Results from lattice simulations}
\label{sec:GWlatticeSimulations}

In this section we present the basic features of the GW spectra produced during the post-inflationary Higgs decay process, obtained from the outcome of our lattice simulations. We leave a detailed parametrization of the spectra for Section~\ref{sec:parametrization}, and the analysis of the redshift of the GW background until today for Section \ref{subsec:GWtoday}. 

Our simulations depend on a series of parameters, some of them being unknown quantities of our system. The first unknown quantity is 
the initial amplitude of the Higgs field {\small$\varphi_*$}. We know that, at the end of inflation, $\varphi_*$ changes from patch to patch with variance given by Eq.~(\ref{eq:VarCaseI}).  Our lattice simulations do correspond to a single patch, inside which we consider the random value {\small$\varphi_*$} to be homogeneous. The second unknown parameter is the amplitude of the Higgs self-coupling {\small$\lambda$} at the post-inflationary scales. This determines the exact form/amplitude of the Higgs potential {\small $V = \lambda\varphi^4/4$} introduced in the lattice. If we fix the strength  of the gauge couplings to their value at very high energies, {\small$g_W^2 \approx 0.3$} and {\small$g_Z^2 \approx 2g_{W} = 0.6$}, the two parameters {\small$(\varphi_*,\lambda)$} can be equivalently replaced by the pair {\small$(\beta,q_s)$}, where {\small$\beta \equiv \lambda^{1/2}\varphi_*/H_*$} [Eq.~(\ref{eq:beta})] characterizes the Higgs amplitude parameter normalized to the unknown inflationary Hubble rate $H_*$, and 
{\small$q_s = (g_Z^2+2g_W^2)/4\lambda$} [Eq.~(\ref{eq:superboson})].

Taking into account the large freedom in the Hubble rate, {\small$10^{2} {\rm GeV} \ll H_* \lesssim 10^{14} {\rm GeV} $}, and the experimental uncertainty in the top quark mass {\small $m_t$} (which affects the running of {\small$\lambda$}), a good physical range for these parameters is {\small$\beta \in [5 \cdot 10^{-4},0.3]$} and {\small$q_s \in [20,3000]$} (corresponding to {\small$\lambda \in [1.5 \cdot 10^{-2},10^{-4}]$}). 

Note that we have also simulated values within the range {\small$5 \leq q_s \leq 20$} for completeness, although for high-energy inflationary scales with {\small$H_*$} of the order of (or somewhat smaller than) {\small $H_*^{\rm max} \sim 10^{14}$} GeV, those values correspond to excessively high {\small$\lambda$}. Only for inflationary Hubble rates {\small$H_* \ll H_*^{\rm (max)}$} we expect to obtain {\small$q_s \lesssim \mathcal{O}(10)$}; however this kills completely the GW signal Eq.~(\ref{eq:TotGW}), as the latter scales as $\propto (H_*/m_p)^4 \ll 1$.

As we do not consider any particular inflationary model, the post-inflationary expansion rate is also unknown. We can characterize it by the equation of state {\small$w$}, see Eq.~(\ref{eq:exprate}). For example, if inflation is caused by an inflaton with a quadratic potential, the Universe following inflation expands effectively as {\rm MD}, with {\small$\rho_{\rm tot} \propto 1/a^3$}. If the inflaton potential is quartic instead, it behaves as {\rm RD} with {\small$\rho_{\rm tot} \propto 1/a^4$}. We can even consider more exotic scenarios, like a {\rm KD} universe, in which the energy density decays faster than that of relativistic species, with {\small$\rho_{\rm tot} \propto 1/a^6$}. As we do not make any assumptions on the particular inflationary model, we consider {\small$w$} as a free parameter determining the expansion rate. In conclusion, we parametrize the GW spectra as a function of three independent variables {\small$q_s$}, {\small$\beta$}, and {\small$w$}.

\begin{figure*}[t]
    \begin{center}
        \includegraphics[width=8cm]{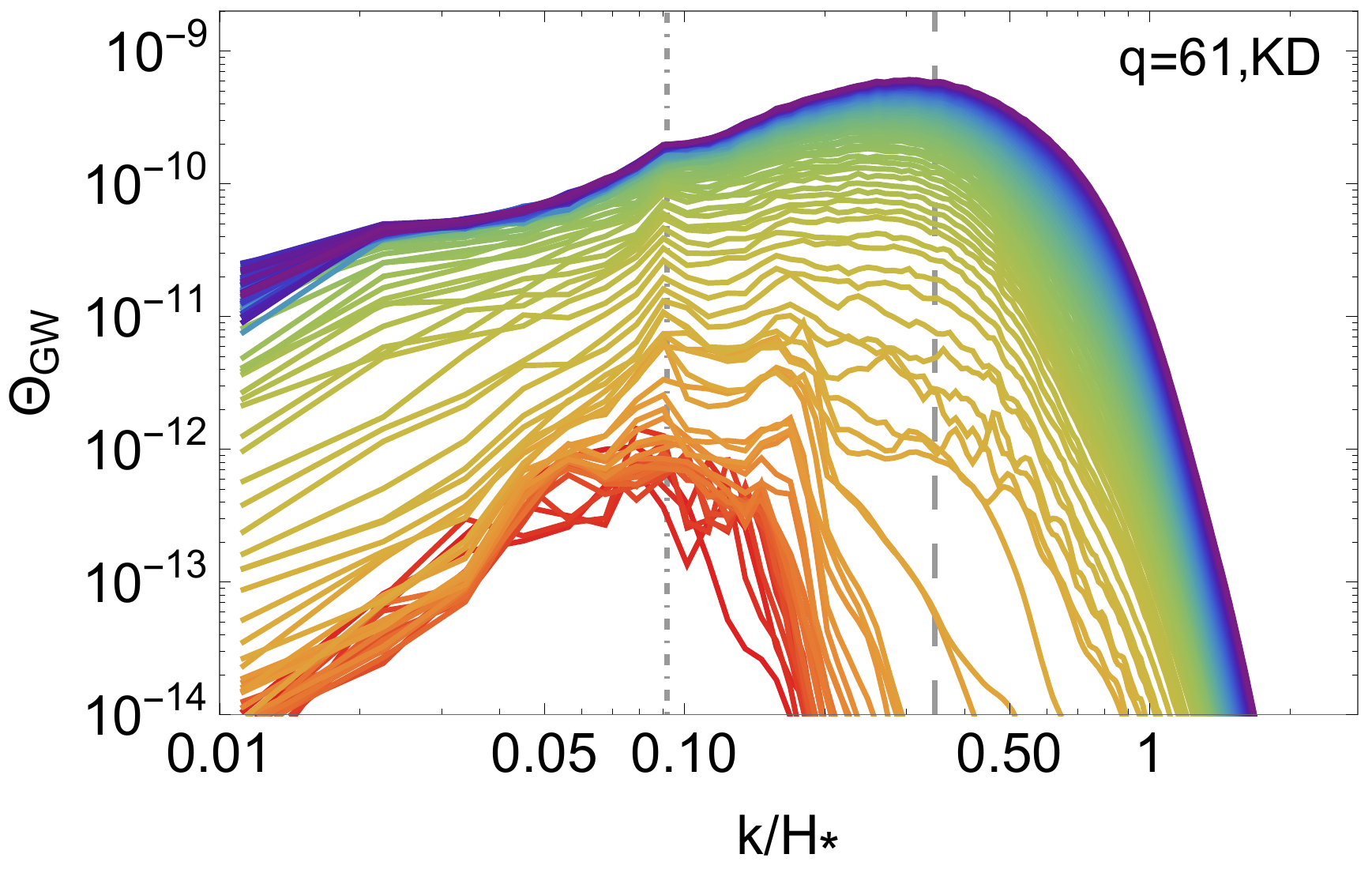}~~~
        \includegraphics[width=8cm]{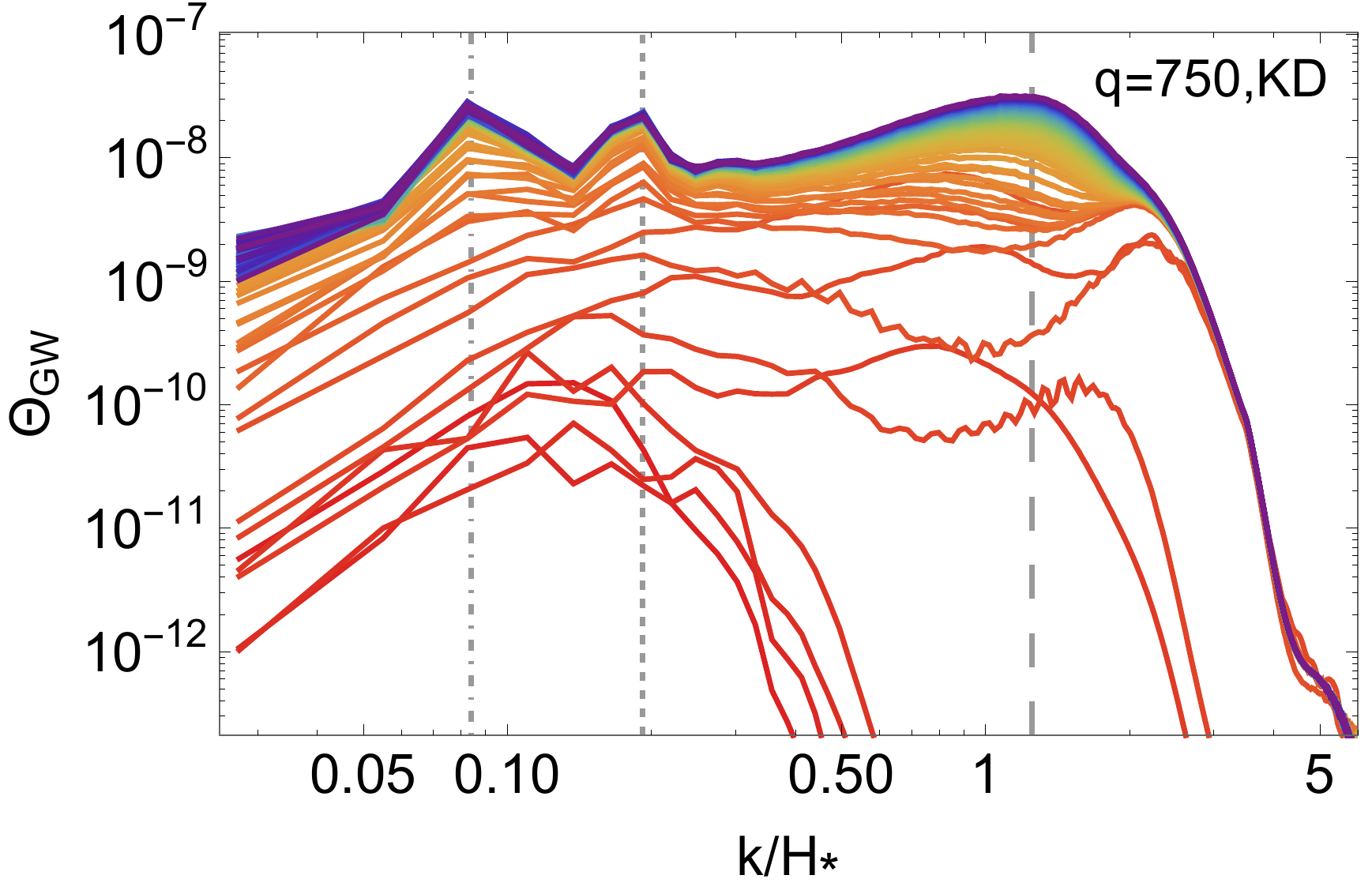}
        \includegraphics[width=8cm]{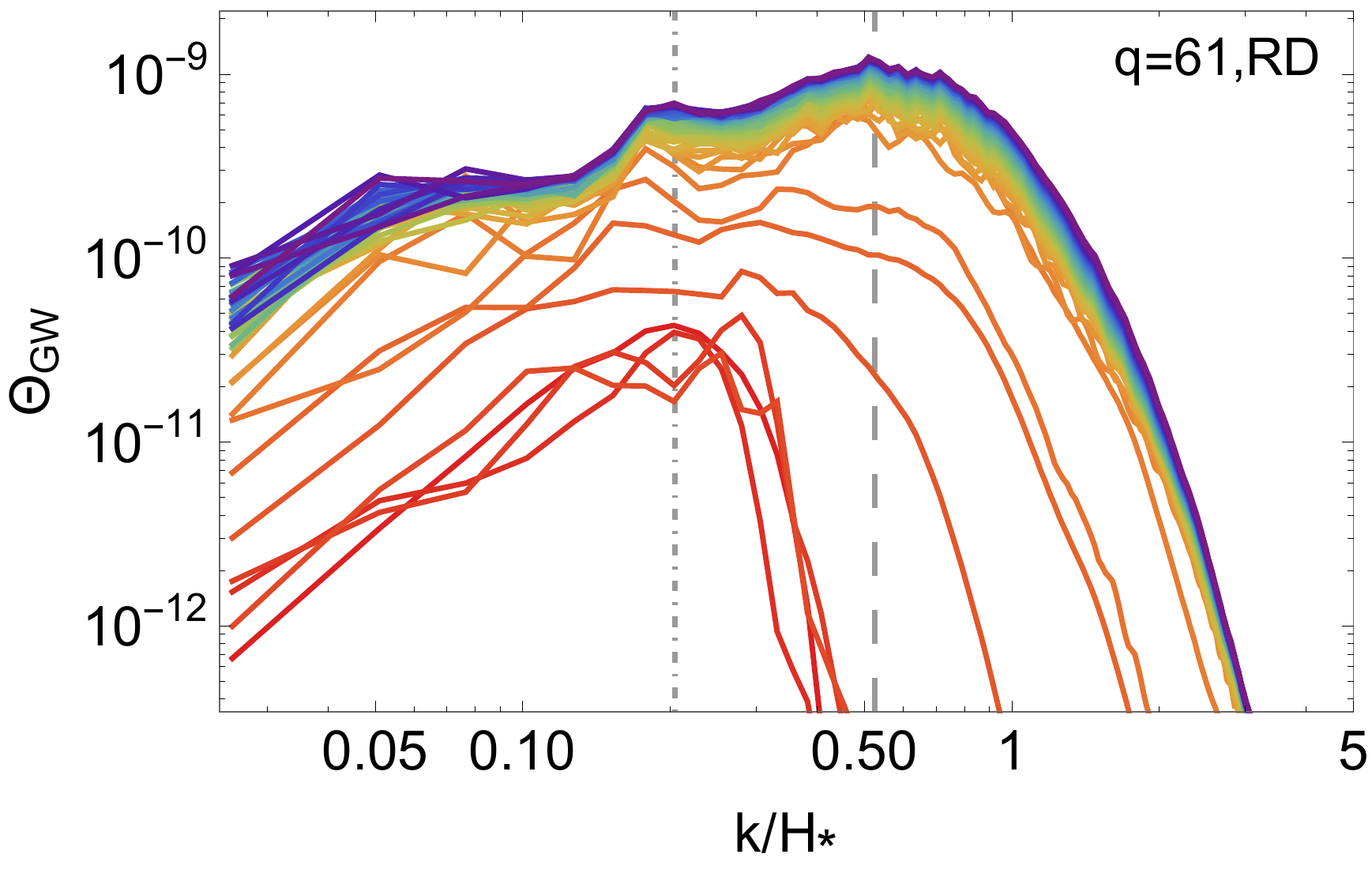}~~~
        \includegraphics[width=8cm]{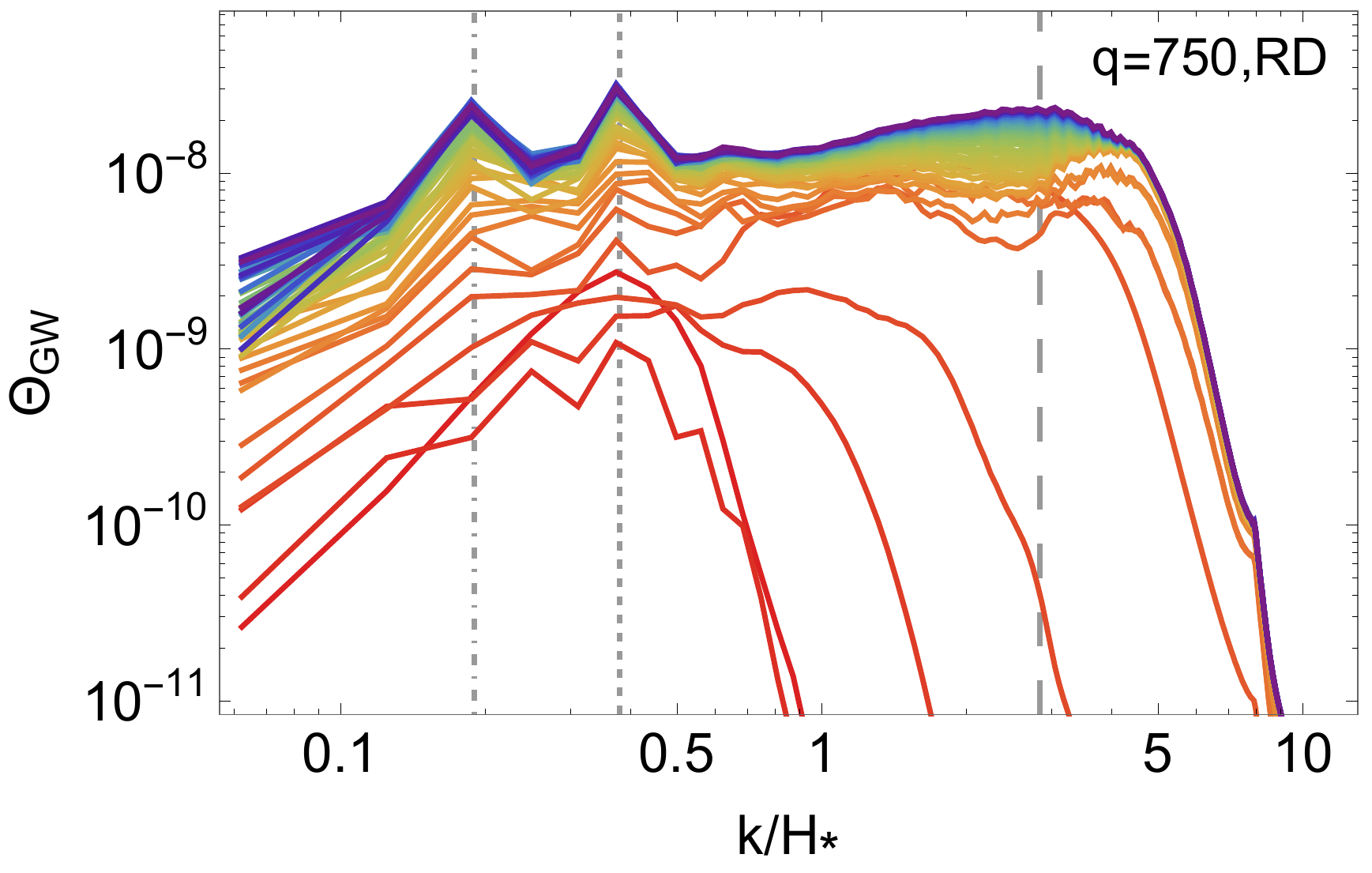}
        \includegraphics[width=8cm]{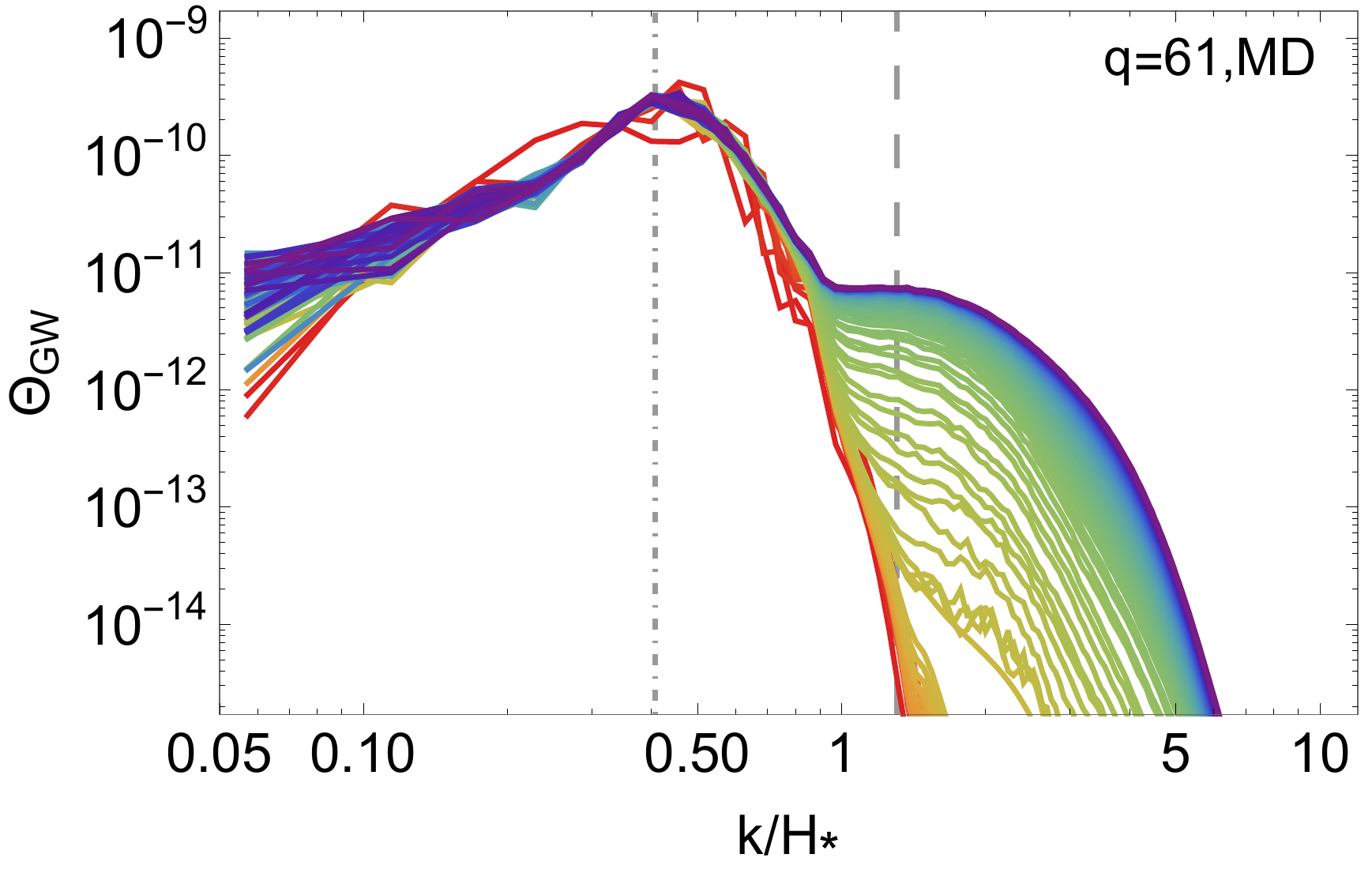}~~~
        \includegraphics[width=8cm]{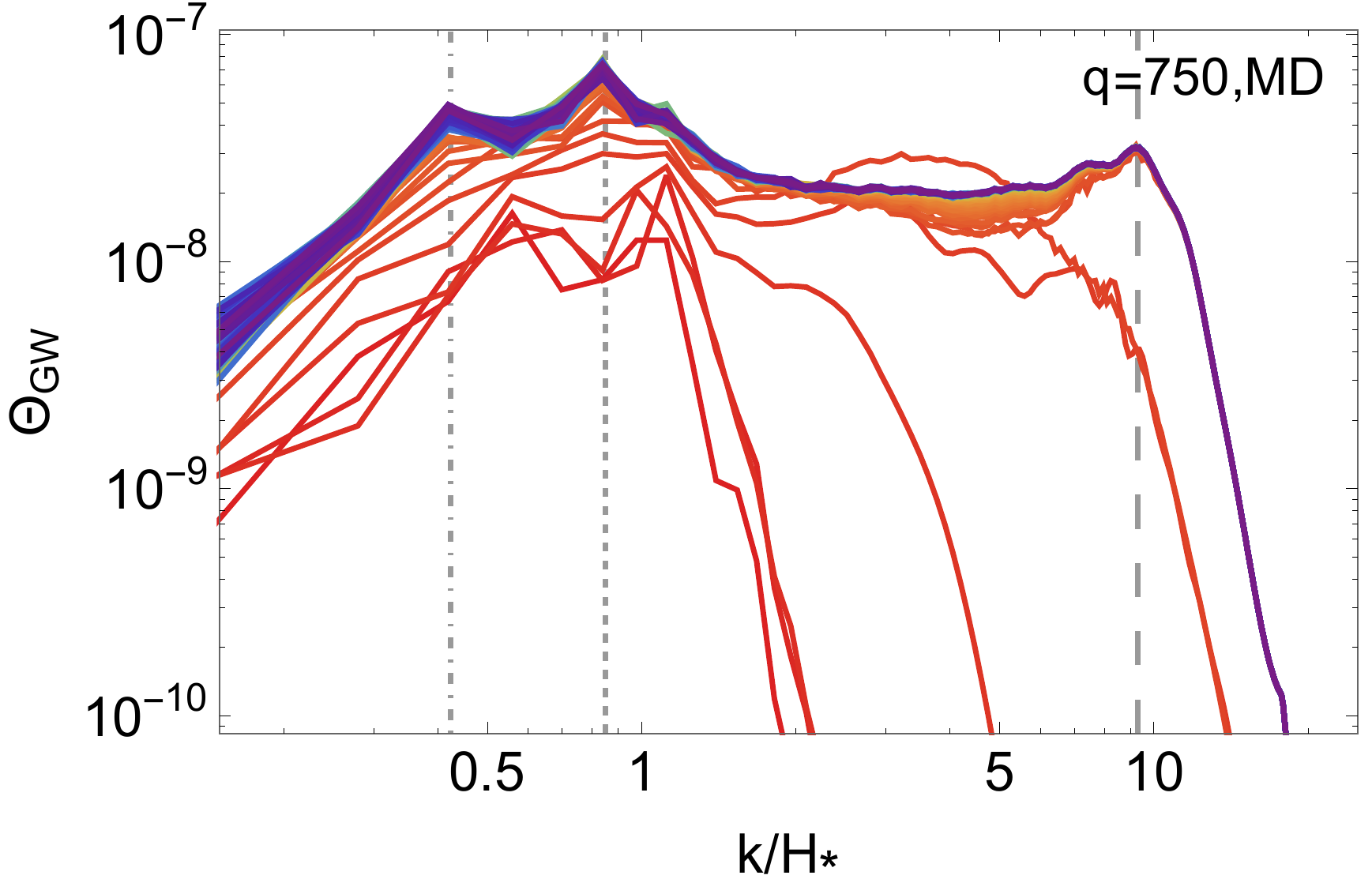}
    \end{center}
\caption{Evolution in time of {\small$\Theta_{{\rm GW}}(k,z;q_s,\beta,w)$} as the GW are being created, computed for the resonance parameters {\small$q_s = 61$} and 750, and {\rm KD}, {\rm RD} and {\rm MD} post-inflationary expansion rates. Within each plot, each colored line corresponds to a particular time, going from early times just after the onset of resonance (red lines) to late times when the growth of the GW ceases (purple lines). The time step between spectra is {\small$\Delta z \approx 32.7$} for KD, {\small$\Delta z \approx 15.5$} for RD, and {\small$\Delta z \approx 7.3$} for MD. The last spectra plotted in each figure corresponds to the output time {\small$z \approx 3280$} in KD, and to {\small$z \approx 750$} in both RD and MD. The dotted-dashed, dashed, and dotted vertical lines indicate the position of various peaks {\small$k_1$}, {\small$k_2$} and {\small$k_3$} in the spectra, see bulk text.} 
    \label{fig:GW-spectra}
\end{figure*}

As described in \cite{Figueroa:2015hda}, the exact dynamics of the Higgs decay process depend sensitively on the value of {\small$q_s$}.  Correspondingly, it is expected that the exact details of the {\rm GW} spectra will also depend sensitively on {\small$q_s$}. However, the qualitative aspects of these spectra can be easily understood, without the need to specify the particular value of {\small$q_s$}. To see this, let us look at Fig.~\ref{fig:GW-spectra}. There we show the temporal evolution of the spectrum {\small$\Theta_{_{\rm GW}}(k,z;q_s,\beta,w)$}. 
The plots correspond to the resonance parameters {\small$q_s = 61$} and {\small$750$}, and to {\rm KD}, {\rm RD} and {\rm MD} post-inflationary expansion rates. Within each plot, each line corresponds to the {\rm GW} spectra at a particular time, showing its evolution from approximately the start of the Higgs oscillations until well after the production of GW ceases. Note that in these plots we consider the particular value {\small$\beta=0.01$}, but a scaling of the results to arbitrary {\small$\beta$} values will be presented in the next section.

Let us now discuss three qualitative aspects of the {\small$\Theta_{_{\rm GW}}(k,z;q_s,\beta,w)$} spectra shown in the figure: the time evolution of the spectra, the amplitude when GW stop growing, and the appearance of peaks.

Let us focus first on the time evolution of the spectra, and its relation with the time scales of the post-inflationary Higgs dynamics introduced in the last section: {\small$z_{\rm osc}$} (onset of the Higgs oscillatory regime), {\small$z_i$} (time at which the backreaction of the gauge bosons onto the Higgs condensate starts becoming effective), and {\small$z_e$} (stabilization of the Higgs energy density and the onset of equipartition). We observe in Fig.~\ref{fig:GW-spectra} that the GW production begins shortly after the start of the Higgs oscillations, i.e.~at the onset of parametric resonance at {\small$z \gtrsim z_{\rm osc}$}. From then on, we observe a significant growth of the {\rm GW} amplitude during the linear stage {\small$z \lesssim z_i$}. This is due to the initial exponential excitation of the gauge bosons, due to the parametric resonance induced by the Higgs condensate oscillations.  However, the final amplitude of the spectra is mostly determined by the non-linear dynamics during some time after the onset of backreaction {\small$z > z_i$}, while the Higgs condensate is decaying noticeably. We can define {\small$z_{_{\rm GW}}$} as the time scale at which GW stop being produced, so that {\small$\Theta_{_{\rm GW}}$} saturates to a fixed amplitude. In general, one finds that {\small$z_{_{\rm GW}} < z_e$}. In other words, the GW stop being produced before the onset of equipartition. This can be clearly observed in Fig.~\ref{fig:GW-spectra}. In Ref.~\cite{Figueroa:2015hda} we found 
$${\small z_e \approx 58.1 \beta^{\frac{-(1 + 3w)}{3 (1 + w)}} q_s^{0.42}}\,,$$
which in the examples shown in the figure, corresponds to {\small$z_e \approx 1520, 3270, 7040$} for {\small$q_s = 67$}, and to {\small$z_e \approx 4350, 9370, 20190$} for {\small$q_s = 750$}, for KD, RD, MD post-inflationary expansion rates, respectively. Note that these times are much longer than the final times displayed in Fig.~\ref{fig:GW-spectra}, when the spectra have already saturated.

The fact that {\small$z_{_{\rm GW}} < z_e$} is indeed not surprising. The precise moment when GW cease to be produced is better determined when the Higgs energy density stops dropping abruptly, and this happens sometime after {\small$z = z_i$} but before {\small$z = z_e$}, see Fig.~\ref{fig:Energy-Components}.  From this time onwards ({\small$z > z_{_{\rm GW}}$}), even if the Higgs energy density is still decaying until the onset of equipartition at $z = z_e$, the matter fields are only evolving smoothly, adjusting themselves towards equipartition. The time {\small$z_e$} simply indicates when the Higgs (comoving) energy density is finally stabilized to a fixed amplitude, coinciding with the onset of equipartition. In conclusion, there is no more GW production after {\small$z = z_{_{\rm GW}}$}. The growth of the {\rm GW} spectra saturates at that moment, and the {\rm GW} simply redshift from then on, due to the expansion of the Universe.

Let us now discuss the final amplitude of {\small$\Theta_{\rm GW}$} after it has saturated, i.e. for {\small$z > z_{_{\rm GW}}$}. If we focus on the panels where {\small$q_s=61$}, we see that, independently of the chosen post-inflationary expansion rate (either KD, RD or MD), the maximum amplitude of the GW spectra is of the same order of magnitude, {\small$\Theta_{\rm GW} \sim \mathcal{O} (10^{-10})$}. Of course, the particular shape of the final spectra is different in each case, but the final amplitude seems to very similar. The same happens if we focus on the {\small$q_s=750$} case, in which, for the three KD, RD and MD spectra, we have {\small$\Theta_{\rm GW} \sim  \mathcal{O} (10^{-8})$}. This indicates that the final amplitude of {\small$\Theta_{\rm GW}$} at saturation is roughly independent on the post-inflationary expansion rate.

This should not be confused, however, with the standard change of amplitude of the GW due to their nature as relativistic species. The prefactor {\small$\epsilon_w$} in Eq.~(\ref{eq:GWfactorsOUT}), which verifies {\small$\epsilon_{w_1} > \epsilon_{w_2}$} if {\small$w_1 > w_2$}, accounts precisely for this effect. Therefore, the final amplitude of the GW is indeed much more affected by their natural redshifting, than by the small dependence of {\small$\Theta_{\rm GW}$} on the rate of expansion. Since {\small$\epsilon_w$} has an exponential dependence on {\small$w$}, see Eq.~(\ref{eq:GWfactorsOUT}), the proportionality {\small$\Omega_{_{\rm GW}} \propto \epsilon_w$} impacts dramatically on the amplitude of the GW background today. It can affect the GW in both ways, either suppressing the amplitude by {\small$\epsilon_{w} < 1$} for {\small$w < 1/3$}, or enhancing it by {\small$\epsilon_{w} > 1$} for {\small$w > 1/3$}. Furthermore, this modulation of the GW amplitude will continue even after the production of GW has ceased, i.e.~at {\small$z > z_{_{\rm GW}}$}. This is because the GW only redshift at the same rate as the expanding background when the Universe becomes RD, hence turning the prefactor into unity, {\small$\epsilon_w = 1$}. In summary, the {\small$\Omega_{_{\rm GW}} \propto \epsilon_{w}$} dependence means that the slower the post-inflationary expansion rate (i.e.~the larger the equation of state {\small$w$}), the higher the final amplitude of the {\rm GW} background.

\begin{figure*}[t]
        \begin{center}
            \includegraphics[width=8cm]{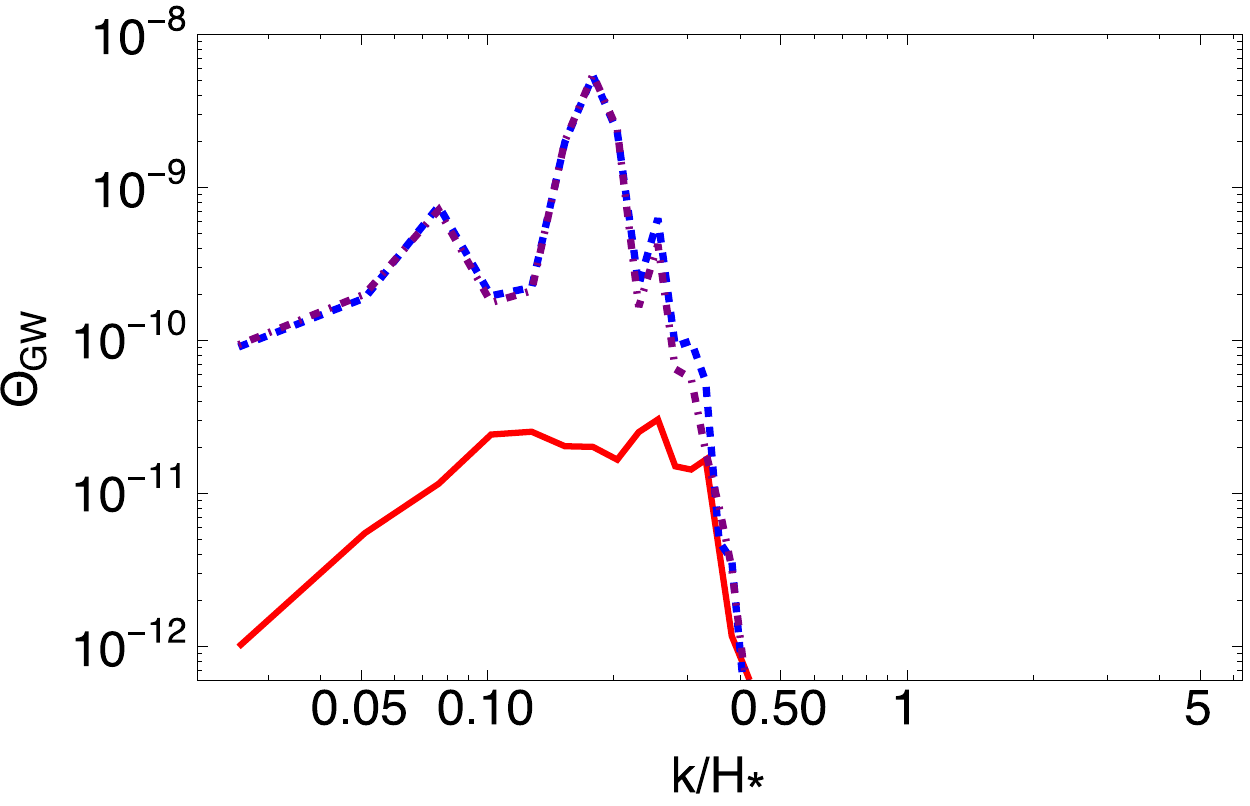}~~~~
            \includegraphics[width=8cm]{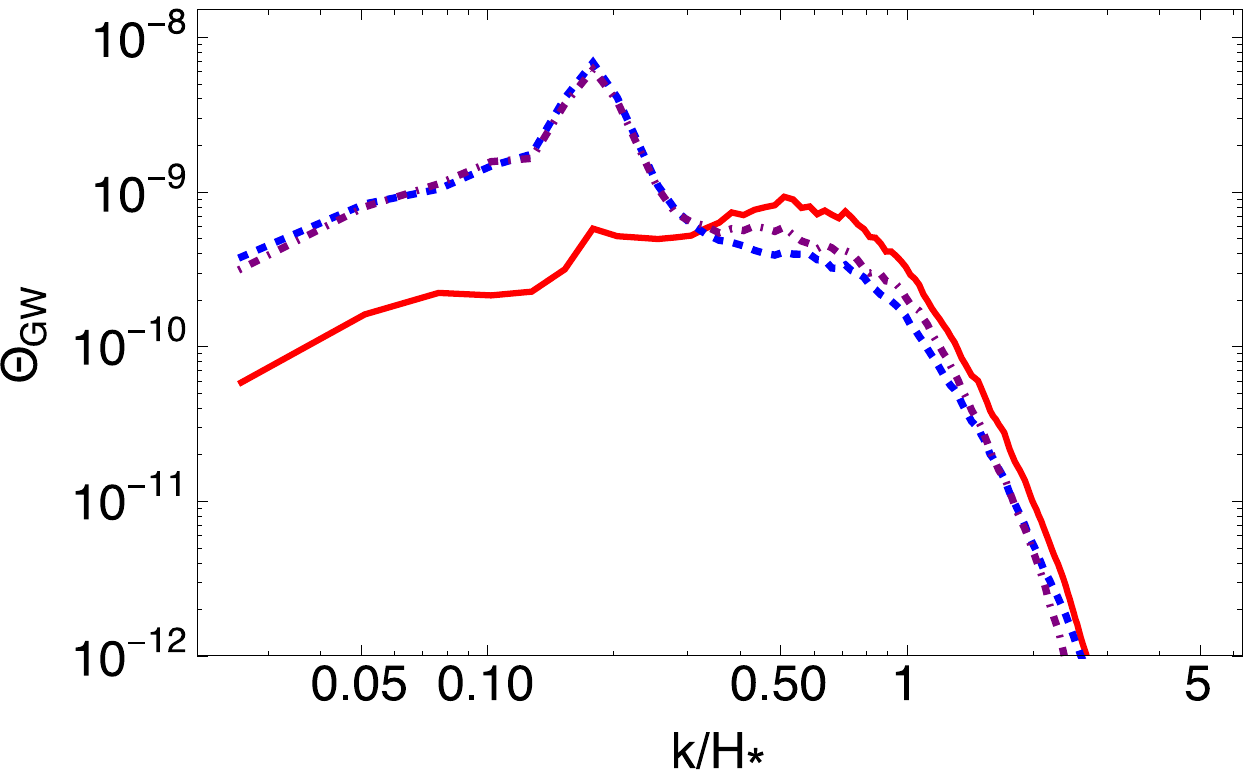}
        \end{center}
        \caption{We show {\small$\Theta_{_{\rm GW}}$} (red continuous line), {\small$\Theta_{_{\rm GW}}^{[h]}$} (dashed blue line) and {\small$\Theta_{_{\rm GW}}^{[g]}$} (dotted-dashed purple line) at the times {\small$z = 62$} (left figure) and {\small$z = 373$} (right figure). The two figures correspond to {\small$q_s = 61$} and {\small$\beta = 0.01$}.} 
        \label{fig:GWinterference}
    \end{figure*} 
    
Let us finally discuss the appearance of peaks in the GW spectra. In Fig.~\ref{fig:GW-spectra} we can see that, during the growth of the GW spectra, a structure of peaks develops. The GW are sourced by both the Higgs and gauge fields through the terms {\small$\mathcal{P}_{ij}$} of Eq.~(\ref{eq:GWsource}), acting in the $rhs$ of Eq.~(\ref{eq:gw-eom}). In momentum space, the spectrum of GW is then sourced by a convolution of the Higgs and gauge fields spectra. Therefore, the position of the peaks is correlated with the appearance of peaks in the spectra of both the Higgs and the gauge fields.

Let us denote by {\small$u_{ij}^{[g]}$} the contribution to the GW sourced only by the gauge fields term {\small$\mathcal{P}_{ij}^{[g]}(\mathcal{E},\mathcal{B})$}, and by {\small$u_{ij}^{[h]}$} the contribution sourced by the Higgs covariant derivatives {\small$\mathcal{P}_{ij}^{[h]}(\mathcal{D}h)$}, see Eq.~(\ref{eq:GWsource}). From the linearity of Eq.~(\ref{eq:GW-nonTTeom}), it follows that
\bea u^{[g]''}_{ij} - \left(\partial_k\partial_k + \frac{a''}{a}\right)u^{[g]}_{ij} &=& \frac{1}{a}\lbrace\mathcal{P}_{ij}^{[g]}\rbrace^{\rm TT} \ , \\
u^{[h]''}_{ij} - \left(\partial_k\partial_k + \frac{a''}{a}\right) u^{[h]}_{ij} &=& \frac{1}{a}\lbrace\mathcal{P}_{ij}^{[h]}\rbrace^{\rm TT} \ . \label{eq:gwGH-eom} 
\eea
Similarly, let us denote as {\small$\Theta_{_{\rm GW}}^{[g]}$} and {\small$\Theta_{_{\rm GW}}^{[h]}$} the contribution to the GW spectra associated to these fields respectively. Clearly, as the GW spectrum is quadratic in {\small$u_{ij}$}, then {\small$\Theta_{_{\rm GW}}$} {\small$ = \Theta_{_{\rm GW}}^{[h]}$} {\small$+~ \Theta_{_{\rm GW}}^{[g]}$} {\small$+~\Theta_{_{\rm GW}}^{[gh]}$}, where {\small$\Theta_{_{\rm GW}}^{[gh]}$} represents an interference contribution from the convolution of a term like {\small$\sim \mathcal{P}_{ij}^{[g]}\mathcal{P}_{ij}^{[h]}$}. In Fig.~\ref{fig:GWinterference} we show, for the case {\small$q_s = 61$}  and  {\small$\beta = 0.01$}, both {\small$\Theta_{_{\rm GW}}^{[g]}$} and {\small$\Theta_{_{\rm GW}}^{[h]}$} as well as the total spectrum {\small$\Theta_{_{\rm GW}}$} for two different times. One can see that {\small$\Theta_{_{\rm GW}}^{[g]}$} and {\small$\Theta_{_{\rm GW}}^{[h]}$} evolve in a similar manner, being almost identical, especially in the infrared regime. In particular, they both show some peaks at certain scales. This is a reflection of the dynamics of the system, which creates similar peaks in the spectra of {\small$\mathcal{E}_i$}, {\small$\mathcal{B}_j$} and {\small$\mathcal{D}_ih$}, correspondingly transferring those peaks into {\small$\mathcal{P}_{ij}^{[g]}$} and {\small$\mathcal{P}_{ij}^{[h]}$}: during the linear regime of parametric resonance, the fast creation of gauge bosons induces a similar growth of the electric and magnetic fields, as well as of the Higgs covariant derivatives. As a consequence, {\small$\mathcal{P}_{ij}^{[g]}$} and {\small$\mathcal{P}_{ij}^{[h]}$} contribute very similarly to the total spectrum of GW. This has in fact a very interesting effect in the {\rm GW} spectrum, as it produces a clear destructive interference effect in the infrared, suppressing the total amplitude {\small$\Theta_{_{\rm GW}}$} with respect the individual amplitudes {\small$ \Theta_{_{\rm GW}}^{[h]}$} {\small$\approx~\Theta_{_{\rm GW}}^{[g]}$}. At the same time, this softens (in some cases it almost washes out) the peak structure, which becomes much more smoothed in the final spectrum. This is clearly shown by the continuous curves in Fig.~\ref{fig:GWinterference}, as compared with the dashed and dotted-dashed curves. 
  
The origin of the peaks can be understood by examining the spectra of the matter fields, i.e.~of the Higgs and gauge bosons\footnote{The interested reader can see such spectra in Section V of~\cite{Figueroa:2015hda}.}. Looking at the initial stages of the process, a growth in both the Higgs and gauge fields spectra takes place in infrared scales (small {\small$k$}). Peaks are generated in the matter fields spectra, according to the band structure of the Lam\'e equation. These peaks are created during the initial stages of the process, when parametric resonance starts building up, 
and the Lam\'e equation applies. These scales are essentially imprinted in the spectrum of the GW during the excitation of the matter fields. The position of the most-infrared peak in the GW spectra, common to both the {\small$q_s = 61$} and 750 cases in Fig.~\ref{fig:GW-spectra}, is indicated with a dotted-dashed line. It corresponds to the initial resonance band in the spectra of the gauge fields. In the {\small$q_s=750$} case, there is even a second peak in the GW spectrum, indicated with a dotted line. It corresponds to another peak appearing in the spectrum of the Higgs field. When the system becomes fully non-linear, the spectra of both fields show a rescattering effect towards the ultraviolet, populating modes of higher and higher momenta. This generates a characteristic feature in the fields' spectra, which develops a relatively wide peak with a 'hunchback' shape in the ultraviolet scales. This last peak is shifted towards higher momenta according to how large the resonance parameter {\small$q_s$} is. Again, this scale is imprinted in the GW spectrum, and it is indicated with a dashed line in both cases {\small$q_s = 61$} and 750 in Fig~\ref{fig:GW-spectra}. \newline

A systematic study of the origin and correlation of the peaks in the GW and the matter fields spectra would be extremely interesting. Such a study could be used to probe the properties of the SM at high energies, such as the running of the couplings involved. This goes, however, beyond the scope of this paper where, as a first step, we only want to obtain a more modest characterization of the general aspects of the GW background. We will therefore adopt a phenomenological approach in section~\ref{sec:parametrization}, characterizing the peak structure in the GW spectra, by means of simple fitting formulas.

\section{Parametrization of the Gravitational Wave Spectra}
\label{sec:parametrization}

In this section we parametrize the position and amplitude of the final peaks in the GW spectra as a function of $q_s$, $w$, and $\beta$. We will focus first on the particular case $\beta=0.01$, and from this, we will apply the scaling found in \cite{Figueroa:2015hda} to extrapolate and generalize this parametrization to other $\beta$ values.

Let us start with the position of the peaks. We show in Figure~\ref{fig:GWfits} the momenta {\small$k_i$} at which the peaks appear as a function of {\small$q_s$}, for the {\small$\beta=0.01$} case, and for the different expansion rates we have simulated: KD, RD and MD. The maximum number of peaks we can observe in the spectra is three: one associated to the hunchback, whose position we denote by {\small$k_3$} (red circles), and two associated with the initial parametric resonance dynamics, whose position we denote by {\small$k_1$} (purple squares) and {\small$k_2$} (orange triangles). However, for some values of $q_s$ we do not see all three peaks: for {\small$q_s \lesssim 200$} the {\small$k_2$} peak is not clearly observed, as it overlaps with either of the two. Also, for some {\small$q_s$} the peaks {\small$k_1$} and {\small$k_3$} are too near to each other, and hence it is difficult to attribute a particular peak to either of them. This explains why, for some specific values of {\small$q_s$} (particularly at low {\small$q_s$}), we just show the red circles corresponding to {\small$k_3$}.

\begin{figure}[t]
    \begin{center}
            \includegraphics[width=7.92cm]{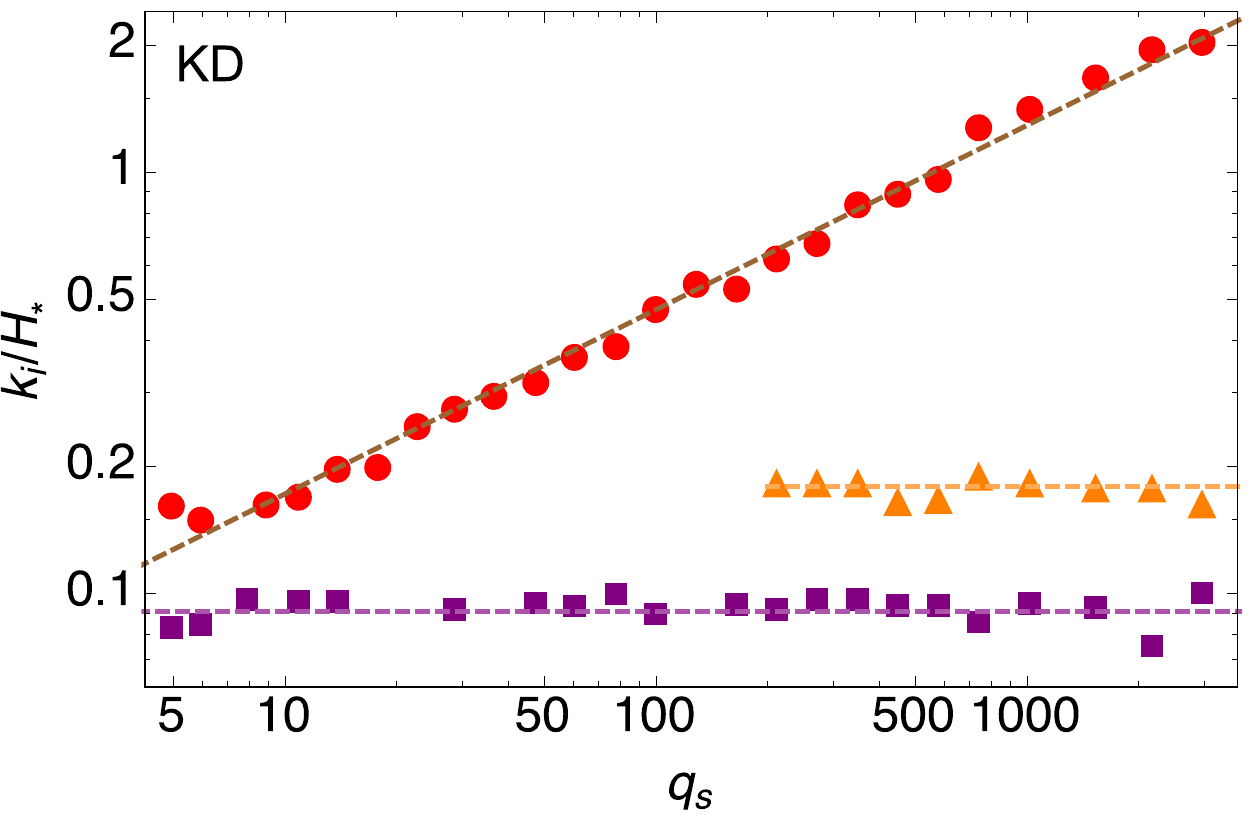}\\
            \includegraphics[width=7.92cm]{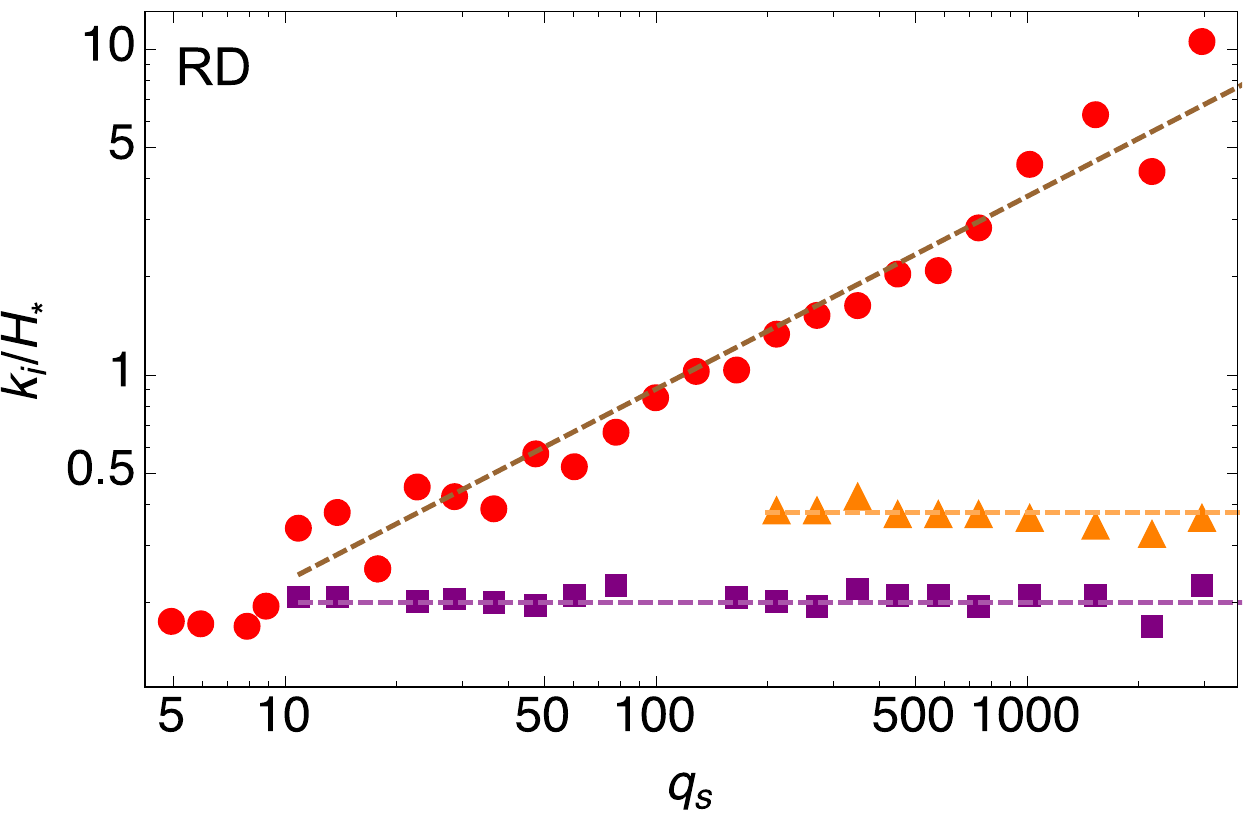}\\
            \includegraphics[width=7.92cm]{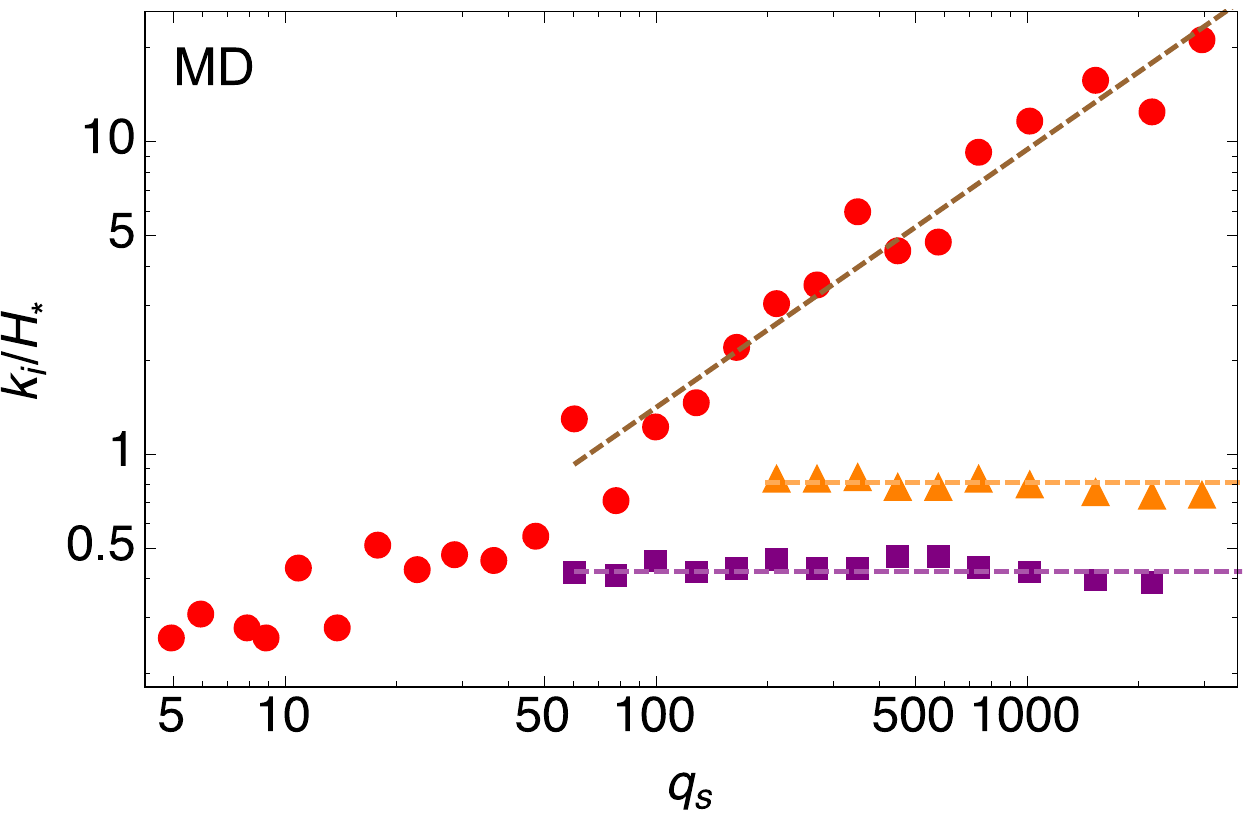}
    \end{center}
\caption{Location of the different peaks {\small$k_i/H_*$} that appear in the GW spectra, as a function of the resonance parameter {\small$q_s$}. The panels correspond to {\small${\rm KD}$} (top), {\small${\rm RD}$} (middle) and {\small${\rm MD}$} (bottom), all obtained for {\small$\beta=0.01$}. Red circles, purple squares and orange triangles correspond to {\small$k_1$}, {\small$k_2$} and {\small$k_3$}, respectively. Dashed lines correspond to the best fits to {\small$k_1$}, {\small$k_2$} and {\small$k_3$}, as given by Eqs.~(\ref{eq:k1fitA})- (\ref{eq:k2k3fit}).} 
        \label{fig:GWfits}
    \end{figure}

The key idea is that, except for very low {\small$q_s$}, we appreciate a clear separation between the hunchback {\small$k_3$} scale and the other scales {\small$k_1, k_2$}. This separation is appreciated in all the post-inflationary expansion rates. More specifically, the position of the hunchback peak increases with {\small$q_s$}, exhibiting a clear power-law dependence.
We find the following fit
\be k_3 \approx A_3 \,q_s^{r} H_* \label{eq:k1fitA}  \ee
with the parameter values (for {\small$\beta=0.01$}) as
\be
A_3\approx \left\{ \begin{array}{ll}
    0.0315, & \mbox{if KD}\\
    0.0593, & \mbox{if RD}\\
    0.0627, & \mbox{if MD}\end{array}  \right. \hspace{0.2cm},\hspace{0.4cm}
r \approx \left\{ \begin{array}{ll}
    0.44, & \mbox{if KD}\\
    0.59, & \mbox{if RD}\\
    0.82, & \mbox{if MD}\end{array} \right. \label{eq:k1fit} \ee

On the other hand, the position of {\small$k_1$} and {\small$k_2$} are mostly independent\footnote{In reality there is a logarithmic dependence to {\small$q_s$}, but we would need to go to very large values {\small$q_s \gg 10^3$} to start noticing it.} on {\small$q_s$}. We find these peaks to be well fitted by
\bea 
k_1 &\approx & A_1  H_*\,,\\
k_2 &\approx & A_2 H_* ~~{\rm [}\,q_s \gtrsim 200\,{\rm]}\,,
\eea
with parameter values (again for {\small$\beta=0.01$}) as
\be
A_{1}\approx \left\{ \begin{array}{ll}
    0.091, & \mbox{if KD}\\
    0.20, & \mbox{if RD}\\
    0.42, & \mbox{if MD}\end{array}  \right.  \hspace{0.1cm},\hspace{0.3cm} 
A_{2}\approx \left\{ \begin{array}{ll}
    0.18, & \mbox{if KD}\\
    0.38 & \mbox{if RD}\\
    0.81, & \mbox{if MD}\end{array}  \right. \label{eq:k2k3fit}
\ee

\noindent These fits are depicted with straight lines in Fig.~\ref{fig:GWfits}.

On the other hand, we show in Fig.~\ref{fig:GWAmplitude}, the amplitude of the spectrum evaluated at the highest peak {\small$\Theta_{_{\rm GW}}(k_p)$}, for the different {\small$q_s$} considered, and for different post-inflationary expansion rates. For {\small$\beta=0.01$}, we find the following phenomenological fit
    \be \Theta_{_{\rm GW}}(k_p) \approx A_{_{\rm GW}} \left(\frac{q_s}{100}\right)^{\alpha_{_{\rm GW}}} \label{eq:Ampfit1} \hspace{0.7cm} (\beta=0.01) \ee
where
    \bea 
    A_{_{\rm GW}} \approx \left\{\hspace*{-1mm}\begin{array}{ll}
        3.1 \times 10^{-9} , & \mbox{if KD}\\
        2.4 \times 10^{-9} , & \mbox{if RD}\\
        2.1 \times 10^{-9} , & \mbox{if MD} \end{array}  \right. \,,\hspace{0.1cm} 
    \alpha_{_{\rm GW}} \approx \left\{\hspace*{-1mm}\begin{array}{ll}
        1.50 , & \mbox{if KD}\\
        1.58  , & \mbox{if RD}\\
        1.61 , & \mbox{if MD} \end{array}  \right.  \label{eq:Ampfit2}\nonumber\\
    \eea
    
    \begin{figure}[t]
\begin{center}
\includegraphics[width=8cm]{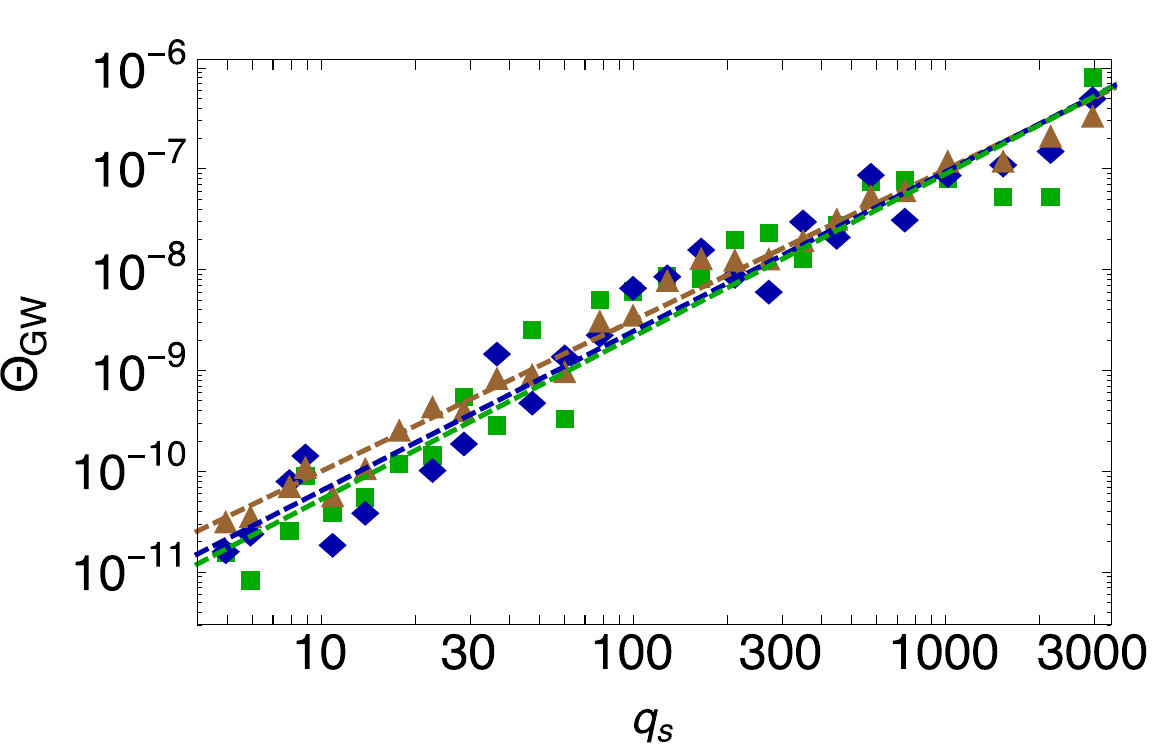}
\end{center}
\caption{Amplitudes {\small$\Theta_{_{\rm GW}}$} of the highest peak of the {\rm GW} spectra as a function of {\small$q$}, and for different post-inflationary expansion rates: KD (brown triangles), RD (blue diamonds) and MD (green squares). Dashed lines correspond to the best-fit functions of Eqs.~(\ref{eq:Ampfit1})-(\ref{eq:Ampfit2}). } 
\label{fig:GWAmplitude}
 \end{figure}
 
This peak corresponds to the maximum amplitude of the GW at the moment when they stop being actively created, i.e., at {\small$z = z_{_{\rm GW}}$}. However, note that {\small$k_p$} does not necessarily correspond always to the same peak {\small$k_1$}, {\small$k_2$} or {\small$k_3$}; rather, it alternates among these [for KD and RD expansion rates we normally have {\small$\Theta_{\rm GW}(k_p) \simeq \Theta_{\rm GW}(k_3)$}, while for MD we have {\small$\Theta_{\rm GW}(k_p) \simeq \Theta_{\rm GW}(k_1)$}]. We see in Fig.~\ref{fig:GWAmplitude} that the three fits for KD, RD, and MD coincide pretty well, confirming what we pointed out in the last section: the maximum amplitude of {\small$\Theta_{\rm GW}$} at saturation time {\small$z_{_{\rm GW}}$} is roughly independent of the post-inflationary expansion rate (the shape, however, is not; see Fig.~\ref{fig:GW-spectra}).

        \begin{figure*}
            \begin{center}
                \includegraphics[width=8cm]{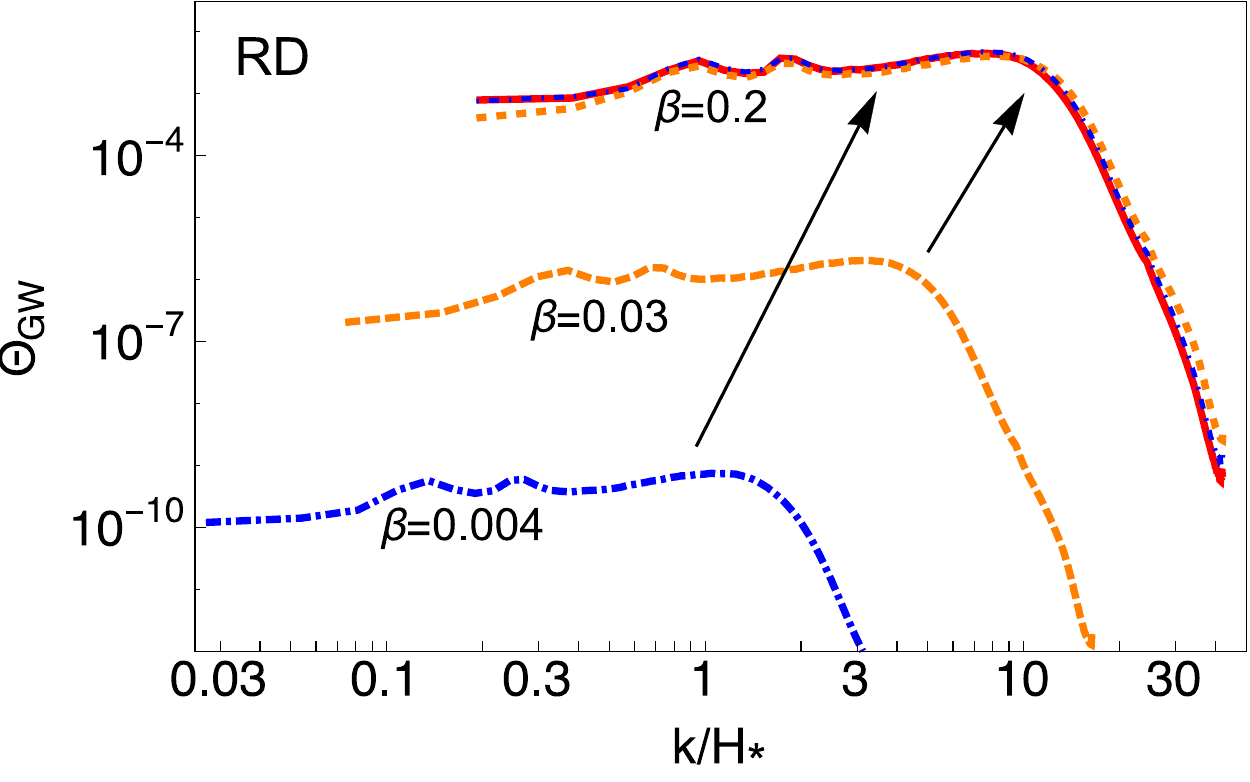}~~~~
                \includegraphics[width=8cm]{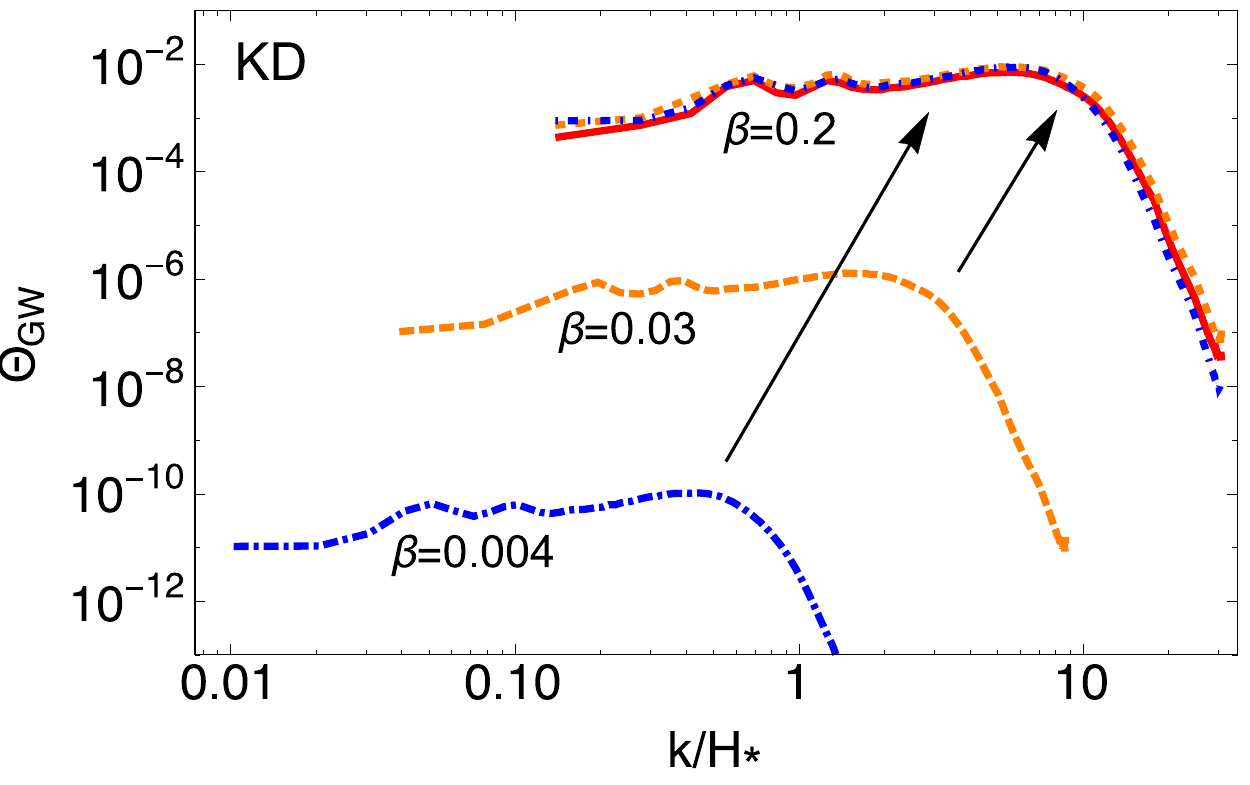}
            \end{center}
                \caption{We show the final spectra {\small$\Theta_{_{\rm GW}}$} for the cases of {\small$\beta=0.2$}, (continuous red line), of {\small$\beta=0.03$} (dashed yellow line), and of {\small$\beta=0.004$} (dot-dashed blue line), obtained directly from lattice simulations. This corresponds to the {\small$q_s = 354$} case, and for RD (left panel) and KD (right panel). We also indicate with arrows the theoretical predictions for the {\small$\beta=0.2$} case, obtained from the {\small$\beta=0.03$} and {\small$\beta=0.004$} lattice results, using the extrapolation laws Eqs.~(\ref{eq:kBetafit}), (\ref{eq:GWbetaScaling}). We can see that the two extrapolated predictions match quite well the output of the real lattice simulations of the {\small$\beta=0.2$} case.} 
            \label{fig:beta-extrap}
        \end{figure*}

These fits have been obtained for the particular {\small$\beta=0.01$} case, but a generic extrapolation to other {\small$\beta$} values can be easily carried out. We just need to use the rescaling laws that we found in~\cite{Figueroa:2015hda}, which connect scales and field amplitudes, from one simulation with Higgs initial amplitude and post-inflationary equation of state {\small$(\beta_1,w_1)$}, to another simulation with different parameters {\small$(\beta_2,w_2)$}. In particular,
\bea
z(\beta_2,\omega_2) &\approx & \beta_1^{p(\omega_1)}\beta_2^{-p(\omega_2)} z(\beta_1,\omega_1)\,,\label{eq:zScaling}\\
k(\beta_2,\omega_2) &\approx & \beta_1^{-p(\omega_1)}\beta_2^{p(\omega_2)} k(\beta_1,\omega_1)\,,\label{eq:kScaling}\\
h(\beta_2,\omega_2) &\approx & \beta_1^{p(\omega_1)-1}\beta_2^{1-p(\omega_2)} z(\beta_1,\omega_1)\,,\label{eq:hScaling}
\eea
where
\be p(w) \equiv \frac{1 + 3 \omega}{3 (1 + \omega)} = \left\{ \begin{array}{ll}
        2/3 , & \mbox{if KD}\\
        1/2, & \mbox{if RD}\\
        1/3, & \mbox{if MD}\end{array} \right. 
    \label{eq:k1fitb}  \ee
    
Using these rescaling laws we predict the position of the peaks in the GW spectrum for arbitrary initial Higgs amplitudes $\beta$ as
\bea  
k_1 &\approx & A_1 \times \left(\frac{\beta}{0.01}\right)^{p(w)} H_* \, \nonumber\\
k_2 &\approx & A_2  \times \left(\frac{\beta}{0.01}\right)^{p(w)} H_* \, \label{eq:kBetafit}\\
k_3 &\approx & A_3  \times \left(\frac{\beta}{0.01}\right)^{p(w)} q_s^{r} H_* \ .\nonumber
\eea

On the other hand, using the scaling laws Eqs.~(\ref{eq:zScaling})-(\ref{eq:hScaling}), we demonstrated in Ref.~\cite{Figueroa:2015hda} that we can recover the dynamics of the matter fields for a given set of {\small$(\beta,w)$} parameters, in terms of the results from an actual simulation done for another set {\small$(\beta',w')$}. Likewise, rescaling the terms involved in the GW source Eq.~(\ref{eq:GWsource}) by means of Eqs.~(\ref{eq:zScaling})-(\ref{eq:hScaling}), we can predict now the scaling of {\small$\Theta_{_{\rm GW}}$} [Eq.~(\ref{eq:thetaGW})], and, hence, how the amplitude of the background of GW scales with {\small$\beta$}. We find that
\be
\Omega_{_{\rm GW}} \propto \Theta_{_{\rm GW}} \propto \beta^{4 + v(w)} \,,\hspace{0.2cm} v(w) =  2\frac{\left(w - {1/3}\right)}{(w + 1)}\,.\label{eq:GWbetaScaling}
\ee

We have confirmed the validity of these predictions by carrying out several lattice simulations with different {\small$\beta$} and {\small$w$} parameters. As an example, in Fig.~\ref{fig:beta-extrap} we show various spectra of GW for {\small$q_s = 354$}, for both RD ({\small$w = 1/3$}) and KD ({\small$w = 1$}). The continuous red, dashed yellow, and dotted-dashed blue lines, show the spectra for {\small$\beta=0.2, 0.03, 0.004$} respectively, obtained directly from lattice simulations. We indicate with arrows the theoretical predictions for {\small$\beta = 0.2$}, as obtained from the {\small$\beta = 0.03$} and {\small$\beta = 0.004$} lattice simulations, using the extrapolation laws Eqs.~(\ref{eq:kBetafit}), (\ref{eq:GWbetaScaling}). We see that the two extrapolated predictions match quite well the output of the real {\small$\beta=0.2$} lattice simulations within errors.

Using Eqs.~(\ref{eq:TotGW}), (\ref{eq:Ampfit1}) and (\ref{eq:GWbetaScaling}), we obtain that the maximum amplitude of the GW background at the end of the production stage, as a function of {\small$\beta, q_s, w$}, is given by
\be \Omega_{_{\rm GW}}(k_p)  \approx A_{_{\rm GW}}\,\delta_{*}\,\epsilon_{w} \left(\frac{q}{100}\right)^{\alpha_{_{\rm GW}}} \left( \frac{\beta}{0.01} \right)^{4 + v(w) } \ , \label{eq:FitFinalAmplitude}\ee
where {\small$\epsilon_w, \delta_{*}$} are given by Eq.~(\ref{eq:GWfactorsOUT}), and {\small$A_{_{\rm GW}}, \alpha_{_{\rm GW}}$} by Eq.~(\ref{eq:Ampfit1}). The amplitude in Eq.~(\ref{eq:FitFinalAmplitude}) constitutes one of the key results of our analysis. However, in order to quantify the amplitude of the signal today, we need to redshift its amplitude and frequency.

\section{The gravitational wave background today: Redshifting the spectrum through cosmic history}
\label{subsec:GWtoday}

We now compute how the GW background redshifts until today. We first define
\be 
\epsilon_{_{\rm RD}} \equiv \epsilon_w(a_{_{\rm RD}}) \equiv \left( \frac{a_*}{a_{_{\rm RD}}} \right)^{(1 - 3 \omega)}\,,
\ee
with {\small$a_*$} the scale factor at the end of inflation at {\small$z = z_* =0$}, {\small$a_{_{\rm RD}}$} the scale factor at the onset of the Radiation-Dominated stage of the Universe at {\small$z = z_{_{\rm RD}}$}, and {\small$w$} the effective equation of state between {\small$z_*$} and {\small$z = z_{_{\rm RD}}$}. Essentially, {\small$\epsilon_{_{\rm RD}}$} quantifies our ignorance about the expansion rate between {\small$z_*$} and {\small$z_{_{\rm RD}}$}. 

Let us take as a frequency of reference the one corresponding to the mode {\small$k_p$} of the highest peak of the spectrum. The frequency today associated to the peak scale {\small$k_p$} is then given by
\bea f_p &\equiv & \left( \frac{a_*}{a_o} \right) \frac{k_p}{2\pi} \\
&=& \epsilon_{_{\rm RD}}^{1/4}\left(\frac{g_{s,\rm{o}}}{g_{s,\rm{RD}}}\right)^{1\over 3}\left(\frac{g_{\rm{o}}}{g_{\rm{RD}}}\right)^{-{1\over 4}}\left(\frac{\rho_{\rm rad}^{\rm (o)}}{\rho_{*}}\right)^{1\over 4}\frac{k_p}{2\pi} \nonumber\\
&\simeq & \epsilon_{_{\rm RD}}^{1/4} \left(\frac{H_*}{H_*^{\rm max}}\right)^{1\over2}\,\frac{k_p}{H_*} \times 2 \cdot 10^{8}~{\rm Hz} \,,
\eea
where {\small$g_{s,t}$} and {\small$g_t$} are the entropic and matter relativistic degrees of freedom at a time {\small$t$}, and we have used the entropy conservation law {\small$a\,T \propto g_{s,t}^{-1/3}$} for a background temperature {\small$T$}, the temperature-energy density relation of a relativistic thermal fluid {\small$\rho \propto g_t\,T^4$}, the evolution of the total energy density of the Universe as {\small$\rho \propto a^{-3(1+w)}$}, the total energy density at the end of inflation {\small$\rho_* = 3m_p^2H_*^2$}, and the value of the energy density of the relativistic species today {\small$\rho_{\rm rad}^{\rm (o)} \approx 2\cdot 10^{-15} eV^4$}. Assuming that the effective degrees of freedom do not change from {\small$z_*$} to {\small$z_{_{\rm RD}}$}, i.e.~{\small$g_{s,*} = g_{s,_{_{\rm RD}}}$} and {\small$g_{*} = g_{_{\rm RD}}$}, and taking into account that {\small$g_{s,t} \sim g_{t}$}, we obtain then {\small$(g_{s,o}/g_{_{s,{\rm RD}}})^{1/3}(g_{o}/g_{_{\rm RD}})^{-{1/4}}$} {\small$\sim (g_{o}/g_{_{\rm RD}})^{1/12} \sim \mathcal{O}(1)$}. 

The amplitude of the GW background today, normalized to the present critical energy density {\small$\rho_c^{(o)}$}, can be written as
\begin{eqnarray}
h^{2}\Omega_{\GW}^{\rm (o)} &\equiv& {h^2\over \rho_{\rm c}^{(o)}}\left(d\rho_{\GW}\over d\log k\right)_{_{\hspace*{-1mm}\rm o}}\\
&=& h^2\Omega_{\rm rad}^{(o)}{1\over\rho_{\rm rad}^{(o)}}\left(a_{_{\rm RD}}\over a_o\right)^4\left(d\rho_{\GW}\over d\log k\right)_{_{\hspace*{-1mm}\rm RD}}\nn
&=& h^2\Omega_{\rm rad}^{\rm (o)}\left(g_{_{s,\rm{o}}}\over g_{_{s,{\rm RD}}}\right)^{4\over3}\left(g_{_{\rm RD}}\over g_{_{\rm o}}\right){1\over\rho_{_{\rm RD}}}\left(d\rho_{\GW}\over d\log k\right)_{_{\hspace*{-1mm}\rm RD}}\nonumber
\end{eqnarray}
where {\small$h^{2}\Omega_{\mathrm{rad}}^{\rm (o)} \equiv h^2\rho_{_{\rm rad}}^{\rm (o)}/\rho_c^{(o)}$}, with {\small$\rho_{_{\rm rad}}^{\rm (o)}$} the radiation component of the Universe today. Using that freely propagating GW scale as a radiation fluid like {\small$\rho_{\GW} \propto 1/a^4$}, {\small$\rho_{_{\rm rad}}^{\rm (o)} = (g_{_{s,{\rm RD}}}/g_{_{s,\rm{o}}})^{4/3}$} {\small$(g_{_{\rm o}}/g_{_{\rm RD}})\rho_{_{\rm RD}}(a_{_{\rm RD}}/a_{_{\rm o}})^4$}, {\small$\rho_{_{\rm RD}} = \rho_*(a_{_{\rm RD}}/a_*)^{-3(1+w)}$} (assuming again that the effective degrees of freedom do not change), and taking into account that {\small$h^{2}\Omega_{\mathrm{rad}}^{\rm (o)} \simeq 4\cdot10^{-5}$} and {\small$(g_{_{s,\rm{o}}}/g_{_{s,{\rm RD}}})^{4/3}(g_{_{\rm RD}}/g_{_{\rm{o}}})$} {\small$\sim (g_{o}/g_{_{\rm RD}})^{1/3} \sim \mathcal{O}(0.1)$}, we can then write the GW energy density spectrum today as
\bea
h^{2}\Omega_{\GW}^{\rm (o)} \simeq \epsilon_{_{\rm RD}}\,\delta_{*} \,\Theta_{_{\rm GW}}\times 10^{-6}\,,
\eea
where {\small$\Theta_{_{\rm GW}}$} [Eq.~(\ref{eq:thetaGW})], read out from the simulations at {\small$z = z_{_{\rm GW}}$}, characterizes the final spectrum shape. 

The highest peak of the GW spectrum today is of course characterized by the highest peak of {\small$\Theta_{_{\rm GW}}$}, parametrized\footnote{Although the highest-amplitude peak {\small$k_p$} is normally {\small$k_3$}, this is not always the case. However, when {\small$k_p$} is instead associated with {\small$k_1$} or {\small$k_2$} (typically for low {\small$q_{s}$}), the spectral amplitude at the {\small$k_3$} peak is still very similar to that of the highest peak. Therefore, for simplicity, we are going to associate here the amplitude {\small$\Theta_{_{\rm GW}}(k_p)$} [Eq.~(\ref{eq:Ampfit1})] to the peak {\small$k_3$} [Eq.~(\ref{eq:k1fit})].} by Eqs.~(\ref{eq:k1fitA}), (\ref{eq:Ampfit1}). 
The frequency and amplitude of highest peak today is then
\bea
f_p \simeq \epsilon_{_{\rm RD}}^{1/4} \left(\frac{H_*}{H_*^{\rm (max)}}\right)^{1\over2}\,\left(\frac{\beta}{0.01}\right)^{p(w)} q_s^{r}~\times~10^{7}~{\rm Hz}\label{eq:FinalFreqPeak}\\
h^{2}\Omega_{\GW}^{\rm (o)}(f_p) \simeq 10^{-24}\times \epsilon_{_{\rm RD}}\,A_{_{\rm GW}}\left(\frac{q_s}{100}\right)^{\alpha_{_{\rm GW}}}\hspace*{0.8cm}\label{eq:FinalAmplitudePeak}\nn
\hspace*{2.8cm}\times \left(\frac{H_*}{H_*^{\rm (max)}}\right)^{4}\left(\frac{\beta}{0.01} \right)^{4+v(w)}
\eea

In order to understand what frequencies and amplitudes these expressions really imply, we need to consider specific cases. For instance, let us assume that the universe is RD after inflation, so that {\small$\epsilon_{_{\rm RD}} = 1$}, and let us consider that the inflationary Hubble rate is close to its upper bound, {\small$H_{*} \lesssim H_*^{\rm (max)}$}. Taking {\small$q_s = 100$} and {\small$\beta_{\rm rms} \simeq 0.1$}, we obtain
\bea
{\rm RD:}~~h^{2}\Omega_{\GW}^{\rm (o)}(f_p) \lesssim 10^{-29}\,,~~{\rm at}~ f_p \lesssim 3 \cdot 10^{8}~{\rm Hz}.\label{eq:FinalPeakRD}
\eea
This amplitude is tiny, so unfortunately there is not much hope to expect to detect it in the future, unless high-frequency GW detection technology undergoes unforeseen development. The main reason why this signal is so small lies in the suppression {\small$\propto \delta_{*} $} {\small$= (H_*/m_p)^4$} {\small$\sim 10^{-18}(H_*/H_*^{\rm (max)})^4 \ll 1$}.

If the Universe was MD after inflation, the situation becomes even worse, because there is an extra dilution of the signal, as the latter is now proportional to some factor {\small$\epsilon_{_{\rm RD}} \ll 1$}, which becomes smaller and smaller the longer it takes for the Universe to reach a RD regime at {\small$z = z_{_{\rm RD}}$}. This dilution is simply a consequence of the fact that GW scale with the expansion of the Universe as relativistic species, {\small$\rho_{_{\rm GW}} \propto 1/a^4$}, whereas a MD background energy density dilutes slower as {\small$\rho \propto 1/a^3$}.

If the Universe is KD after inflation, the GW signal is, however, enhanced significantly. In particular, given the initial ratio of energies {\small$\Delta \equiv V_*/\rho_* \sim 10^{-12}$} [Eq.~(\ref{eq:InitialRatioEnergies})], the Universe will sustain a KD expansion rate until the moment when the relativistic SM fields become dominating the energy budget. This implies that the GW signal is enhanced by a factor {\small$\propto \epsilon_{_{\rm RD}} = 1/\Delta \sim 10^{12}$}. The scaling of the signal also goes as {\small$\propto (\beta/0.01)^{4+v(1)}$} with {\small$v(1) = 2/3$}, instead of with {\small$v(1/3) = 0$} as in RD. In addition, {\small$A_{_{\rm GW}}^{\rm KD} \gtrsim A_{_{\rm GW}}^{\rm RD}$}. Compared to a RD background, and for {\small$\beta = 0.1$}, there is therefore another enhancement (however milder) by a factor {\small$(A_{_{\rm GW}}^{\rm KD}/A_{_{\rm GW}}^{\rm RD})(0.1/0.01)^{v(1)-v(1/3)} \sim 10$}. Plugging all this into Eqs.~(\ref{eq:FinalFreqPeak}),(\ref{eq:FinalAmplitudePeak}), we obtain
\bea
{\rm KD:}~~h^{2}\Omega_{\GW}^{\rm (o)}(f_p) \lesssim 10^{-16}\,,~~{\rm at}~ f_p \lesssim 3 \cdot 10^{11}~{\rm Hz}.\label{eq:FinalPeakKD}
\eea
This corresponds yet to a small signal, but its amplitude is in fact comparable\footnote{In reality, the comparison to the inflationary signal is not fair here, as the KD regime after inflation would also boost the amplitude of the inflationary background by a factor {\small$\propto \epsilon_{_{\rm RD}} \sim 10^{12}$}.} to the standard scale-invariant inflationary background {\small$h^2\Omega_{_{\rm GW}}^{\rm (Inf)} \simeq 5\cdot 10^{-16}(H_*/H_*^{\rm (max)})^2$}. Our signal in this case of KD after inflation, however, lies at extremely high frequencies {\small$\sim 10^{11}$} Hz, beyond the range of planned GW detectors.

Before we move into the concluding section, it is perhaps interesting to make a remark about a particular aspect of this GW background, which we have not studied in this paper. Given the condition of the Higgs as a condensate with a finite correlation length at the end of inflation, it is indeed expected that this GW background will have anisotropies on angular scales corresponding to that correlation scale. This is very similar to the case of preheating scenarios with light preheat fields~\cite{Bethke:2013aba,Bethke:2013vca}. Of course, given the smallness of the background amplitude itself, detecting such anisotropic variation in the sky seems even more chimeric than detecting the background itself. Yet, it is interesting to note that these anisotropies are expected in the present case, as opposed to the situation when the decaying (oscillatory) field is an inflaton, which is considered to be homogeneous\footnote{Of course, the inflaton is not completely homogeneous, as it has fluctuations in order to explain the primordial density perturbations. However, those fluctuations are very tiny compared to its zero mode, whereas in the case at hand of the SM Higgs as a spectator field, the Higgs amplitude varies substantially from patch to patch (of size of its correlation length), with respect its averaged amplitude.} over scales much larger than our present horizon.

\section{Summary and Discussion}
\label{sec:Discussion}

Independently of the nature of the inflationary sector, a stochastic background of GW is expected simply due to the existence of the Standard Model Higgs. This background of GW is always generated after inflation, as long as the Higgs is decoupled from (or sufficiently weakly coupled to) the inflationary sector. In such a case, the Higgs is excited either: $i)$ during inflation, if it is minimally coupled to gravity (or if it has a weak non-minimal coupling), or $ii)$ at the end of inflation, if it has a (sufficiently) large non-minimal coupling to gravity. Either way, we expect the Higgs to be in the form of a condensate after inflation, decaying very rapidly -- via non-perturbative effects -- into the rest of the SM species. The resulting post-inflationary out-of-equilibrium dynamics of the SM fields, generates necessarily a stochastic background of GW. Both the SM Higgs and the electroweak gauge bosons act as the dominant sources of the GW background.

We have studied the details of the form of the {\rm GW} spectrum, determining its frequency, amplitude and shape. We have characterized the {\rm GW} spectrum dependence on the unknown parameters of the system, namely the Higgs initial amplitude at the end of inflation {\small$\beta = \sqrt{\lambda}\varphi_*/H_*$}, the equation of state {\small$w$} characterizing the post-inflationary expansion rate of the Universe, and the resonance parameter {\small$q_s = (g_Z^2+2g_W^2)/4\lambda$}. The running of the Higgs self-coupling at high energies is in fact quite uncertain within the experimental input, so {\small$\lambda$} can vary within the range {\small$10^{-2} \lesssim \lambda < 10^{-5}$}. This translates into some uncertainty in the regime of the resonance parameter, which may vary within the range {\small$q_s \sim \mathcal{O}(10)-\mathcal{O}(10^3)$}. 

We have used real-time classical gauge field lattice simulations in an expanding box in {\small$(3+1)$} dimensions. We chose {\small$N = 256$} points per dimension, ensuring that the relevant modes involved in this process were well captured within the dynamical range of the simulations. Our results have been obtained within an Abelian-Higgs modeling, expected to describe sufficiently well the system when {\small$q_s \gg 1$}. Only in the case of the smallest resonance parameters {\small$q_s \sim \mathcal{O}(10)$}, does one expect the dynamics of the system to be affected by the presence of the full non-Abelian structure of the SM. 

From our lattice simulations, we have obtained Eq.~(\ref{eq:FitFinalAmplitude}), which is a phenomenological fit of the amplitude of the GW spectra as a function of the different unknown parameters described above. We also obtain a parametrization of the observed redshifted amplitude until today in Eq.~(\ref{eq:FinalAmplitudePeak}). However, the GW signal is suppressed by the inflationary Hubble rate as {\small$\propto (H_*/m_p)^4$}. The largest amplitudes for the GW background are therefore obtained when {\small$H_*$} is only somewhat smaller than (but of the order of) its current upper bound {\small$H_*^{\rm (max)} \sim 10^{14}$ GeV}. This implies that {\small$\lambda$} runs to small values {\small$\lambda < 10^{-2}$}, hence making the resonance parameter large, {\small$q_s > 10$}. In light of this, the use of the Abelian approach is fully justified. In any case, the basic features of the fields dynamics and GW production, i.e.~its dependence on {\small$q_s$}, {\small$\beta$} and {\small$w$}, are not expected to change drastically in the full non-Abelian scenario. Our study can be considered therefore as a good indicator of the GW amplitudes to expect in general, even if non-Abelian corrections were to be considered.

If the Universe was RD after inflation, our calculations show in fact that this background is tiny, with an amplitude of {\small$h^{2}\Omega_{\GW}^{\rm (o)}(f_p) \lesssim 10^{-29}$}, and peaked at high frequencies {\small$f_p \sim 300~{\rm MHz}$}. The smallness of this background reflects simply the fact that the initial energy of the Higgs condensate represents only a tiny fraction of the inflationary energy. If the Universe was MD after inflation, although the background will be peaked at slightly smaller frequencies, its amplitude today can only be even smaller than in the RD case. The amplitude of the background is expected, however, to be enhanced significantly if the Universe underwent a KD regime after inflation. In that case, our calculations show that the background today could have an amplitude up to {\small$h^{2}\Omega_{\GW}^{\rm (o)}(f_p) \lesssim 10^{-16}$}. This larger background is, however, peaked at very high frequencies, of the order of {\small$f_p \lesssim  10^{11}~{\rm Hz}$}. 

The generation of the GW background we have studied in this paper is, in some sense, universally expected, as long as the SM is not strongly coupled to the inflationary sector. However, given that the background is always peaked at very high frequencies, and its amplitude today is very small (in all cases), our prediction will remain, most likely, as a curiosity of the SM. 

\acknowledgments
F.T. thanks UNIGE and CERN Theory Division for kind hospitality. J.G.B. thanks the CERN Theory Division for its kind hospitality during the summer of 2015, when this work was developed. This work is supported by the Research Project of the Spanish MINECO, Grant No. FPA2013-47986-03-3P, and the Centro de Excelencia Severo Ochoa Program No. SEV-2012-0249. F.T. is supported by the FPI-Severo Ochoa Ph.D. fellowship No. SVP-2013-067697. We acknowledge the use of the IFT Hydra cluster for the development of this work.

\bibliography{HiggsReheating8}

\begin{thebibliography}{10}

\bibitem{HulseTaylor:1975a}
R.~Hulse and J.~Taylor,
\newblock Astrophys. J. {\bf 195}, L51 (1975).

\bibitem{GWdetection:abc}
B.~A. et~al. [LIGO~Scientific and V.~Collaboration],
\newblock Phys.Rev.Lett. {\bf 116}, 061102 (2016), [1602.03837],
\newblock 10.1103/PhysRevLett.116.061102.

\bibitem{StarobinskyGW1979}
A.~A. Starobinsky,
\newblock JETP Lett. {\bf 30}, 682 (1979).

\bibitem{GWs1997KhlebnikovTkachev}
S.~Khlebnikov and I.~Tkachev,
\newblock Phys.Rev. {\bf D56}, 653 (1997), [hep-ph/9701423],
\newblock 10.1103/PhysRevD.56.653.

\bibitem{GarciaBellido:1998gm}
J.~Garcia-Bellido,
\newblock {Preheating the universe in hybrid inflation},
\newblock in {\em {Proceedings, 33rd Rencontres de Moriond fundamental
  parameters in cosmology}}, pp. 29--34, 1998, [hep-ph/9804205].

\bibitem{GWs2006EastherLim}
R.~Easther and E.~A. Lim,
\newblock JCAP {\bf 0604}, 010 (2006), [astro-ph/0601617],
\newblock 10.1088/1475-7516/2006/04/010.

\bibitem{GWsPRL2007FigueroaGarciaBellido}
J.~Garcia-Bellido and D.~G. Figueroa,
\newblock Phys.Rev.Lett. {\bf 98}, 061302 (2007), [astro-ph/0701014],
\newblock 10.1103/PhysRevLett.98.061302.

\bibitem{GWsPRD2007FigueroaGarciaBellidoSastre}
J.~Garcia-Bellido, D.~G. Figueroa and A.~Sastre,
\newblock Phys.Rev. {\bf D77}, 043517 (2008), [0707.0839],
\newblock 10.1103/PhysRevD.77.043517.

\bibitem{GWs2007DufauxEtAl}
J.~F. Dufaux, A.~Bergman, G.~N. Felder, L.~Kofman and J.-P. Uzan,
\newblock Phys.Rev. {\bf D76}, 123517 (2007), [0707.0875],
\newblock 10.1103/PhysRevD.76.123517.

\bibitem{GWs2009HybridInfDufauxEtAl}
J.-F. Dufaux, G.~Felder, L.~Kofman and O.~Navros,
\newblock JCAP {\bf 0903}, 001 (2009), [0812.2917],
\newblock 10.1088/1475-7516/2009/03/001.

\bibitem{Kamionkowski:1993fg}
M.~Kamionkowski, A.~Kosowsky and M.~S. Turner,
\newblock Phys.Rev. {\bf D49}, 2837 (1994), [astro-ph/9310044],
\newblock 10.1103/PhysRevD.49.2837.

\bibitem{Caprini:2007xq}
C.~Caprini, R.~Durrer and G.~Servant,
\newblock Phys.Rev. {\bf D77}, 124015 (2008), [0711.2593],
\newblock 10.1103/PhysRevD.77.124015.

\bibitem{Huber:2008hg}
S.~J. Huber and T.~Konstandin,
\newblock JCAP {\bf 0809}, 022 (2008), [0806.1828],
\newblock 10.1088/1475-7516/2008/09/022.

\bibitem{Hindmarsh:2013xza}
M.~Hindmarsh, S.~J. Huber, K.~Rummukainen and D.~J. Weir,
\newblock Phys.Rev.Lett. {\bf 112}, 041301 (2014), [1304.2433],
\newblock 10.1103/PhysRevLett.112.041301.

\bibitem{Hindmarsh:2015qta}
M.~Hindmarsh, S.~J. Huber, K.~Rummukainen and D.~J. Weir,
\newblock Phys. Rev. {\bf D92}, 123009 (2015), [1504.03291],
\newblock 10.1103/PhysRevD.92.123009.

\bibitem{GWsAbelianHiggsHybridPreheatingDufauxEtAl2010}
J.-F. Dufaux, D.~G. Figueroa and J.~Garcia-Bellido,
\newblock Phys.Rev. {\bf D82}, 083518 (2010), [1006.0217],
\newblock 10.1103/PhysRevD.82.083518.

\bibitem{GWsGeneral2004DamourVilenkin}
T.~Damour and A.~Vilenkin,
\newblock Phys.Rev. {\bf D71}, 063510 (2005), [hep-th/0410222],
\newblock 10.1103/PhysRevD.71.063510.

\bibitem{Vachaspati:1984gt}
T.~Vachaspati and A.~Vilenkin,
\newblock Phys.Rev. {\bf D31}, 3052 (1985),
\newblock 10.1103/PhysRevD.31.3052.

\bibitem{JonesSmith:2007ne}
K.~Jones-Smith, L.~M. Krauss and H.~Mathur,
\newblock Phys.Rev.Lett. {\bf 100}, 131302 (2008), [0712.0778],
\newblock 10.1103/PhysRevLett.100.131302.

\bibitem{GWsSelfOrderingScalarFieldsFenuEtAl2009}
E.~Fenu, D.~G. Figueroa, R.~Durrer and J.~Garcia-Bellido,
\newblock JCAP {\bf 0910}, 005 (2009), [0908.0425],
\newblock 10.1088/1475-7516/2009/10/005.

\bibitem{GWsUniversalBackgroundFromCDsFigueroaEtAl2012}
D.~G. Figueroa, M.~Hindmarsh and J.~Urrestilla,
\newblock Phys.Rev.Lett. {\bf 110}, 101302 (2013), [1212.5458],
\newblock 10.1103/PhysRevLett.110.101302.

\bibitem{Maggiore2000}
M.~Maggiore,
\newblock Phys. Rept. {\bf 331}, 283 (2000), [gr-qc/9909001],
\newblock 10.1016/S0370-1573(99)00102-7.

\bibitem{Acernese:2015gua}
Virgo, F.~Acernese,
\newblock J. Phys. Conf. Ser. {\bf 610}, 012014 (2015),
\newblock 10.1088/1742-6596/610/1/012014.

\bibitem{Harry:2010zz}
LIGO Scientific, G.~M. Harry,
\newblock Class. Quant. Grav. {\bf 27}, 084006 (2010),
\newblock 10.1088/0264-9381/27/8/084006.

\bibitem{Somiya:2011np}
KAGRA, K.~Somiya,
\newblock Class. Quant. Grav. {\bf 29}, 124007 (2012), [1111.7185],
\newblock 10.1088/0264-9381/29/12/124007.

\bibitem{AmaroSeoane:2012km}
P.~Amaro-Seoane {\em et~al.},
\newblock 1201.3621.

\bibitem{ATLAS2012}
ATLAS Collaboration, G.~Aad {\em et~al.},
\newblock Phys.Lett. {\bf B716}, 1 (2012), [1207.7214],
\newblock 10.1016/j.physletb.2012.08.020.

\bibitem{CMS2012}
CMS Collaboration, S.~Chatrchyan {\em et~al.},
\newblock Phys.Lett. {\bf B716}, 30 (2012), [1207.7235],
\newblock 10.1016/j.physletb.2012.08.021.

\bibitem{Espinosa:2007qp}
J.~Espinosa, G.~Giudice and A.~Riotto,
\newblock JCAP {\bf 0805}, 002 (2008), [0710.2484],
\newblock 10.1088/1475-7516/2008/05/002.

\bibitem{Enqvist:2013kaa}
K.~Enqvist, T.~Meriniemi and S.~Nurmi,
\newblock JCAP {\bf 1310}, 057 (2013), [1306.4511],
\newblock 10.1088/1475-7516/2013/10/057.

\bibitem{Figueroa:2014aya}
D.~G. Figueroa,
\newblock JHEP {\bf 1411}, 145 (2014), [1402.1345],
\newblock 10.1007/JHEP11(2014)145.

\bibitem{KariEtAlNonAbelian14}
K.~Enqvist, S.~Nurmi and S.~Rusak,
\newblock JCAP {\bf 1410}, 064 (2014), [1404.3631],
\newblock 10.1088/1475-7516/2014/10/064.

\bibitem{Figueroa:2015hda}
D.~G. Figueroa, J.~Garcia-Bellido and F.~Torrenti,
\newblock Phys.Rev. {\bf D92}, 083511 (2015), [1504.04600],
\newblock 10.1103/PhysRevD.92.083511.

\bibitem{Enqvist:2015nrw}
K.~Enqvist, S.~Nurmi, S.~Rusak and D.~Weir,
\newblock JCAP {\bf 1602}, 057 (2016), [1506.06895],
\newblock 10.1088/1475-7516/2016/02/057.

\bibitem{Enqvist:2014bua}
K.~Enqvist, T.~Meriniemi and S.~Nurmi,
\newblock JCAP {\bf 1407}, 025 (2014), [1404.3699],
\newblock 10.1088/1475-7516/2014/07/025.

\bibitem{Hook:2014uia}
A.~Hook, J.~Kearney, B.~Shakya and K.~M. Zurek,
\newblock JHEP {\bf 1501}, 061 (2015), [1404.5953],
\newblock 10.1007/JHEP01(2015)061.

\bibitem{Kobakhidze:2013tn}
A.~Kobakhidze and A.~Spencer-Smith,
\newblock Phys.Lett. {\bf B722}, 130 (2013), [1301.2846],
\newblock 10.1016/j.physletb.2013.04.013.

\bibitem{Kobakhidze:2014xda}
A.~Kobakhidze and A.~Spencer-Smith,
\newblock 1404.4709.

\bibitem{Spencer-Smith:2014woa}
A.~Spencer-Smith,
\newblock 1405.1975.

\bibitem{Shkerin:2015exa}
A.~Shkerin and S.~Sibiryakov,
\newblock Phys.Lett. {\bf B746}, 257 (2015), [1503.02586],
\newblock 10.1016/j.physletb.2015.05.012.

\bibitem{Espinosa:2015qea}
J.~R. Espinosa {\em et~al.},
\newblock JHEP {\bf 09}, 174 (2015), [1505.04825],
\newblock 10.1007/JHEP09(2015)174.

\bibitem{Gross:2015bea}
C.~Gross, O.~Lebedev and M.~Zatta,
\newblock Phys.Lett. {\bf B753}, 178 (2015), [1506.05106],
\newblock 10.1016/j.physletb.2015.12.014.

\bibitem{Branchina:2015nda}
V.~Branchina and E.~Messina,
\newblock 1507.08812.

\bibitem{DiVita:2015bha}
S.~Di~Vita and C.~Germani,
\newblock Phys.Rev. {\bf D93}, 045005 (2015), [1508.04777],
\newblock 10.1103/PhysRevD.93.045005.

\bibitem{DiLuzio:2015iua}
L.~Di~Luzio, G.~Isidori and G.~Ridolfi,
\newblock Phys.Lett. {\bf B753}, 150 (2015), [1509.05028],
\newblock 10.1016/j.physletb.2015.12.009.

\bibitem{Grobov:2015ooa}
A.~V. Grobov, R.~V. Konoplich and S.~G. Rubin,
\newblock Annalen Phys.  (2015),
\newblock 10.1002/andp.201500127.

\bibitem{Kamada:2015aqa}
K.~Kamada,
\newblock Phys. Lett. {\bf B742}, 126 (2015), [1409.5078],
\newblock 10.1016/j.physletb.2015.01.024.

\bibitem{Bezrukov:2007ep}
F.~L. Bezrukov and M.~Shaposhnikov,
\newblock Phys.Lett. {\bf B659}, 703 (2008), [0710.3755],
\newblock 10.1016/j.physletb.2007.11.072.

\bibitem{Bezrukov:2010jz}
F.~Bezrukov, A.~Magnin, M.~Shaposhnikov and S.~Sibiryakov,
\newblock JHEP {\bf 1101}, 016 (2011), [1008.5157],
\newblock 10.1007/JHEP01(2011)016.

\bibitem{Starobinsky:1994bd}
A.~A. Starobinsky and J.~Yokoyama,
\newblock Phys.Rev. {\bf D50}, 6357 (1994), [astro-ph/9407016],
\newblock 10.1103/PhysRevD.50.6357.

\bibitem{Kunimitsu:2012xx}
T.~Kunimitsu and J.~Yokoyama,
\newblock Phys. Rev. {\bf D86}, 083541 (2012), [1208.2316],
\newblock 10.1103/PhysRevD.86.083541.

\bibitem{Herranen:2014cua}
M.~Herranen, T.~Markkanen, S.~Nurmi and A.~Rajantie,
\newblock Phys.Rev.Lett. {\bf 113}, 211102 (2014), [1407.3141],
\newblock 10.1103/PhysRevLett.113.211102.

\bibitem{Herranen:2015ima}
M.~Herranen, T.~Markkanen, S.~Nurmi and A.~Rajantie,
\newblock Phys. Rev. Lett. {\bf 115}, 241301 (2015), [1506.04065],
\newblock 10.1103/PhysRevLett.115.241301.

\bibitem{VaccaHiggs}
R.~Casadio, P.~Iafelice and G.~Vacca,
\newblock Nucl.Phys. {\bf B783}, 1 (2007), [hep-th/0702175],
\newblock 10.1016/j.nuclphysb.2007.05.015.

\bibitem{Bezrukov:2008ut}
F.~Bezrukov, D.~Gorbunov and M.~Shaposhnikov,
\newblock JCAP {\bf 0906}, 029 (2009), [0812.3622],
\newblock 10.1088/1475-7516/2009/06/029.

\bibitem{GarciaBellido:2008ab}
J.~Garcia-Bellido, D.~G. Figueroa and J.~Rubio,
\newblock Phys.Rev. {\bf D79}, 063531 (2009), [0812.4624],
\newblock 10.1103/PhysRevD.79.063531.

\bibitem{Figueroa:2009jw}
D.~G. Figueroa,
\newblock AIP Conf.Proc. {\bf 1241}, 578 (2010), [0911.1465],
\newblock 10.1063/1.3462688.

\bibitem{Enqvist:2012im}
K.~Enqvist, D.~G. Figueroa and T.~Meriniemi,
\newblock Phys.Rev. {\bf D86}, 061301 (2012), [1203.4943],
\newblock 10.1103/PhysRevD.86.061301.

\bibitem{Figueroa:2013vif}
D.~G. Figueroa and T.~Meriniemi,
\newblock JHEP {\bf 1310}, 101 (2013), [1306.6911],
\newblock 10.1007/JHEP10(2013)101.

\bibitem{Greene:1998nh}
P.~B. Greene and L.~Kofman,
\newblock Phys.Lett. {\bf B448}, 6 (1999), [hep-ph/9807339],
\newblock 10.1016/S0370-2693(99)00020-9.

\bibitem{Greene:2000ew}
P.~B. Greene and L.~Kofman,
\newblock Phys.Rev. {\bf D62}, 123516 (2000), [hep-ph/0003018],
\newblock 10.1103/PhysRevD.62.123516.

\bibitem{Degrassi:2012ry}
G.~Degrassi {\em et~al.},
\newblock JHEP {\bf 1208}, 098 (2012), [1205.6497],
\newblock 10.1007/JHEP08(2012)098.

\bibitem{Bezrukov:2012sa}
F.~Bezrukov, M.~Y. Kalmykov, B.~A. Kniehl and M.~Shaposhnikov,
\newblock JHEP {\bf 1210}, 140 (2012), [1205.2893],
\newblock 10.1007/JHEP10(2012)140.

\bibitem{Ade:2015xua}
Planck, P.~A.~R. Ade {\em et~al.},
\newblock 1502.01589.

\bibitem{Casas:1994qy}
J.~A. Casas, J.~R. Espinosa and M.~Quiros,
\newblock Phys. Lett. {\bf B342}, 171 (1995), [hep-ph/9409458],
\newblock 10.1016/0370-2693(94)01404-Z.

\bibitem{Casas:1996aq}
J.~A. Casas, J.~R. Espinosa and M.~Quiros,
\newblock Phys. Lett. {\bf B382}, 374 (1996), [hep-ph/9603227],
\newblock 10.1016/0370-2693(96)00682-X.

\bibitem{DeSimone:2012qr}
A.~De~Simone and A.~Riotto,
\newblock JCAP {\bf 1302}, 014 (2013), [1208.1344],
\newblock 10.1088/1475-7516/2013/02/014.

\bibitem{ATLAS:2014wva}
ATLAS, CDF, CMS, D0,
\newblock 1403.4427.

\bibitem{Ballesteros:2015iua}
G.~Ballesteros and C.~Tamarit,
\newblock JHEP {\bf 1509}, 210 (2015), [1505.07476],
\newblock 10.1007/JHEP09(2015)210.

\bibitem{Greene:1997fu}
P.~B. Greene, L.~Kofman, A.~D. Linde and A.~A. Starobinsky,
\newblock Phys.Rev. {\bf D56}, 6175 (1997), [hep-ph/9705347],
\newblock 10.1103/PhysRevD.56.6175.

\bibitem{GarciaBellido:2003wd}
J.~Garcia-Bellido, M.~Garcia-Perez and A.~Gonzalez-Arroyo,
\newblock Phys.Rev. {\bf D69}, 023504 (2004), [hep-ph/0304285],
\newblock 10.1103/PhysRevD.69.023504.

\bibitem{GarciaBellido:2001cb}
J.~Garcia-Bellido and E.~Ruiz~Morales,
\newblock Phys.Lett. {\bf B536}, 193 (2002), [hep-ph/0109230],
\newblock 10.1016/S0370-2693(02)01820-8.

\bibitem{DaniPhD}
D.~G. Figueroa,
\newblock {\em Aspects of Reheating},
\newblock PhD thesis, 2010.

\bibitem{Figueroa:GWformalism}
D.~G. Figueroa, J.~Garcia-Bellido and A.~Rajantie,
\newblock JCAP {\bf 1111}, 015 (2011), [1110.0337],
\newblock 10.1088/1475-7516/2011/11/015.

\bibitem{Bethke:2013aba}
L.~Bethke, D.~G. Figueroa and A.~Rajantie,
\newblock Phys.Rev.Lett. {\bf 111}, 011301 (2013), [1304.2657],
\newblock 10.1103/PhysRevLett.111.011301.

\bibitem{Bethke:2013vca}
L.~Bethke, D.~G. Figueroa and A.~Rajantie,
\newblock JCAP {\bf 1406}, 047 (2014), [1309.1148],
\newblock 10.1088/1475-7516/2014/06/047.

\end{thebibliography}
\bibliographystyle{h-physrev4}

\end{document}